\documentclass[twocolumn,twocolappendix]{aastex63}

\newcommand\ds{{\sc Dark Sage}}
\newcommand\HI{H\,{\sc i}}

\received{Feb 1, 2022}
\revised{May 10, 2022}
\accepted{\today}
\submitjournal{ApJ}

\def\be{\begin{equation}}
\def\ee{\end{equation}}
\def\CF3{{\sc cosmicflows-3}}
\usepackage{rotating}
\usepackage{multirow}
\usepackage{CJKutf8}
\usepackage{graphicx}	
\usepackage{amsmath}	
\usepackage{amssymb}	

\usepackage{newtxtext,newtxmath}

\usepackage[T1]{fontenc}
\usepackage{ae,aecompl}

\usepackage{color}

\shorttitle{HOD of H\thinspace{\protect\scriptsize I}  Galaxies}
\shortauthors{Qin et al.}
\graphicspath{{./}{figures/}}

\begin{document}

\title{H\thinspace{\protect\scriptsize I} HOD -- I. The Halo Occupation Distribution of H\thinspace{\protect\scriptsize I}  Galaxies}

\correspondingauthor{Fei Qin}
\email{feiqin@kasi.re.kr}

\author{Fei Qin}
\altaffiliation{Korea Astronomy and Space Science Institute, Yuseong-gu, Daedeok-daero 776, Daejeon 34055, Republic of Korea}
\affiliation{Korea Astronomy and Space Science Institute, Yuseong-gu, Daedeok-daero 776, Daejeon 34055, Republic of Korea}

\author{Cullan Howlett}
\affiliation{School of Mathematics and Physics, The University of Queensland, Brisbane, QLD 4072, Australia}

\author{Adam R.~H.~Stevens}
\affiliation{International Centre for Radio Astronomy Research, The University of Western Australia, Crawley, WA 6009, Australia}

\author{David Parkinson}
\affiliation{Korea Astronomy and Space Science Institute, Yuseong-gu, Daedeok-daero 776, Daejeon 34055, Republic of Korea}



\begin{abstract}

The next generation of galaxy surveys will provide more precise measurements of galaxy clustering than have previously been possible. The 21-cm radio signals that are emitted from  neutral atomic hydrogen (H\thinspace{\protect\scriptsize I}) gas will be detected by large-area radio surveys such as WALLABY and the SKA, and deliver galaxy positions and velocities that can be used to measure galaxy clustering statistics. But, to harness this information to improve our cosmological understanding, and learn about the physics of dark matter and dark energy, we need to accurately model the manner in which galaxies detected in H\thinspace{\protect\scriptsize I} trace the underlying matter distribution of the Universe. For this purpose, we develop a new H\thinspace{\protect\scriptsize I}-based Halo Occupation Distribution (HOD)  model, which makes predictions for the number of galaxies present in dark matter halos conditional on their H\thinspace{\protect\scriptsize I} mass.  The parameterised HOD model is fit and validated using the {\sc Dark Sage} semi-analytic model, where we show that the HOD parameters can be modelled by simple linear and quadratic functions of H\thinspace{\protect\scriptsize I} mass. However, we also find that the clustering predicted by the HOD depends sensitively on the radial distributions of the H\thinspace{\protect\scriptsize I} galaxies within their host dark matter halos, which does not follow the NFW profile in the {\sc Dark Sage} simulation. As such, this work enables -- for the first time -- a simple prescription for placing galaxies of different H\thinspace{\protect\scriptsize I} mass within dark matter halos in a way that is able to reproduce the H\thinspace{\protect\scriptsize I} mass-dependent galaxy clustering and H\thinspace{\protect\scriptsize I} mass function simultaneously   and without requiring knowledge of the optical properties of the galaxies. Further efforts are required to demonstrate that this model can be used to produce large ensembles of mock galaxy catalogues for upcoming surveys.

\end{abstract}



\section{Introduction} \label{sec:intro}

The large-scale structure of matter in the Universe, as traced out by the distribution of galaxies and light, provides an important observational probe of the physics of the origin and evolution of the Universe. The clustering statistics of the three-dimensional galaxy distribution can be used to measure the expansion rate of the Universe (through Baryon Acoustic Oscillations) and the rate at which structures form (using redshift-space distortions and peculiar velocities). These cosmological probes have already been used as  powerful tools to test the standard cosmological model of $\Lambda$ cold dark matter ($\Lambda$CDM), and constrain the parameters relating to dark matter and dark energy \citep{Gorski1989,feldman1994,Hatton1998,Cole2005,Yamamoto2006,Beutler2012,Johnson2014,Carrick2015,Blake2018,Shi2018,Dupuy2019,Howlett2019,Qin2019b,Qin2019a}. As the next generation of surveys increase in depth, volume and population number, these cosmological measurements will only increase in precision.

However, guaranteeing the accuracy of such clustering measurements requires detailed modelling of the galaxy survey. Accurate cosmological inferences need to include the manner in which the observed ``tracer galaxies'' map to the underlying dark matter distribution, the effects of the survey selection function, geometry and completeness, and the possible ensemble of clustering measurements (of which our Universe is only a single realisation) known as cosmic variance.

These effects are most commonly accounted for using large ensembles of numerical simulations, for example, the {\sc WizCOLA} \citep{10.1093/mnras/stw763}, L-PICOLA \citep{Howlett2015}, Quijote \citep{VillaescusaNavarro2019}, or AbacusSummit \citep{Maksimova2021} simulation suites. A popular method for then populating these simulations with galaxies is the Halo Occupation Distribution (HOD, \citealt{Jing1998,Peacock2000,Berlind2002,Zheng2004,Tinker2005,Guo2013} and references therein). These methods do not track the formation of a galaxy from gas in the simulation directly. Instead the HOD is a statistical approach, using an empirical relationship between dark matter halo mass and the probability of finding galaxies in that halo at a given position or with certain properties. This  semi-empirical relationship comes from either simulations that involve galaxy formation, or else from existing data sets.

If one can directly use galaxy formation theory \& simulations to generate fake galaxies and fake surveys, why use the HOD approach? This is mainly due to the fact that it is slower to generate galaxies using galaxy formation theory, and requires robust and accurate dark matter halo properties and merger trees. For the large ensembles of simulations required to estimate the variance in clustering measurements, the mass resolution is typically limited and the halo properties often only reproduced approximately (see i.e., \citealt{Manera2015, Howlett2015,Blot2019}). However, using an HOD, one can quickly generate a large number, usually thousands, of mock galaxy surveys and compensate for the lack of fidelity in the underlying dark matter and halo clustering.

To date, most cosmological measurements from galaxy surveys have relied on optical redshifts to map the large-scale structure in our Universe. However, in the coming years, we will start to see the advent of large-scale radio surveys being used for precision cosmology. For example the Widefield ASKAP L-band Legacy All-sky Blind surveY (or WALLABY, \citealt{Koribalski2020}) and the Square Kilometre Array (SKA, \citealt{2020PASA...37....7S}) will provide a large number of late-type galaxies with observed redshifts, and distances measured using the Tully--Fisher relation (which is the empirical relation between a galaxy's absolute magnitude and its H\thinspace{\protect\scriptsize I} rotation width; \citealt{Tully1977,Strauss1995,Masters2008,Hong2019,Kourkchi2020}). To harness those surveys for cosmology, we need to understand how galaxies detected in H\thinspace{\protect\scriptsize I} trace the  underlying matter distribution of the Universe. This essentially demands the development of an HOD conditional on the H\thinspace{\protect\scriptsize I} mass. This is important, as not every 21-cm galaxy (or H\thinspace{\protect\scriptsize I}{\color{red}-}emitting object) will populate the dark matter halo in the same way, regardless of mass. Instead the relationship between H\thinspace{\protect\scriptsize I} mass and dark matter halo mass will be complicated by astrophysical effects during galaxy formation.  An H\thinspace{\protect\scriptsize I}-conditional HOD allows us to build up a realistic model of the radio galaxy distribution, not just by position on the sky, but also by H\thinspace{\protect\scriptsize I} mass. This is analogous to other work that has focused on conditional luminosity and stellar mass functions (e.g., \citealt{Cooray2006, Guo2018}).

The relation between H\thinspace{\protect\scriptsize I} mass and dark matter halo mass has been explored in many previous works \citep{Padmanabhan2017,Castorina2017,Villaescusa2018,Obuljen2019,Guo2020,Chauhan2020,Avila2022}.  For example, \citet{Obuljen2019} and \citet{Guo2020} constrain the parameters in the H\thinspace{\protect\scriptsize I}–halo mass relation using the ALFALFA survey \citep{Giovanelli2005}.
\citet{Chauhan2020} studied the H\thinspace{\protect\scriptsize I}–halo mass relation using the semi-analytic galaxy formation model {\sc SHARK} \citep{Lagos2018,Lagos2018_2}, where they explore the effect of different physical processes on the shape of the H\thinspace{\protect\scriptsize I}–halo mass relation. However, these cases have only explored one aspect of the problem at a time: either considering the total H\thinspace{\protect\scriptsize I} mass in a given halo, or the number of H\thinspace{\protect\scriptsize I} galaxies across a survey (perhaps as a function of environment, but largely irrespective of halo mass).     \citealt{Paranjape2021,Paranjape2021b} using  an optical HOD of \citealt{Guo2015b} and scaling relation of \citealt{Paul2018} to predict the clustering of H\thinspace{\protect\scriptsize I} selected  galaxies in different H\thinspace{\protect\scriptsize I} mass bins, assuming the optical and H\thinspace{\protect\scriptsize I} selected samples statistically trace out the same underlying halo distribution. However, this requires knowledge of the optical properties of the galaxies.  

In this work, we model both    aspects  simultaneously    without requiring knowledge of the optical properties of the galaxies , and develop a methodology for producing simulations for 21-cm radio galaxy (hereafter, H\thinspace{\protect\scriptsize I} galaxy) surveys such as the WALLABY and SKA, which will measure the galaxy position and redshift through neutral hydrogen emission. In this first paper, we develop an improved conditional HOD model; in later work we will demonstrate how this can be painted onto approximate simulations before applying survey selection effects.

The paper is structured as follows: In Section \ref{sec:datasim}, we introduce the {\sc Dark Sage} simulation, which will be used to explore the connection between H\thinspace{\protect\scriptsize I} mass, halo mass and galaxy clustering. In Section \ref{sec:hod}, we introduce the H\thinspace{\protect\scriptsize I}-mass conditional HOD that we will be fitting. In Section \ref{sec:Xigg}, we introduce the galaxy two-point correlation function, and the halo properties that are used to model it.  In Sections \ref{sec:app} and \ref{sec:Result} we apply this modelling to the {\sc Dark Sage} simulation, showing how the parameters of the model can be fit first from the galaxy distribution as a function of halo mass, and then from only the galaxy clustering. We conclude in Section \ref{sec:conc}.
Hereafter, the little letter $m$ denotes the mass in real-space, while the capital letter $M \equiv \log_{10}m/m_{\odot}$ denotes the mass in log-space per unit solar mass $m_{\odot}$.

\section{The D{\small ARK} S{\small AGE} model} \label{sec:datasim}

In this work we use a simulated catalogue of galaxies to fit an \HI-based HOD that is conditional on \HI~mass (i.e. it makes different predictions for the halo distribution of that object based on its neutral hydrogen mass). This simulated catalogue is built from the {\sc Dark Sage} semi-analytic model of galaxy formation, specifically the version of \citet{Stevens2018}.

Originally presented in \citet{Stevens2016}\,---\,with updates in \citet{Stevens2017,Stevens2018}\,---\,\ds~builds and evolves galaxies within halo merger trees constructed from a cosmological $N$-body simulation.  The version used here was run on the standard merger trees of the Millennium simulation \citep{Springel2005}, widely used throughout the literature.  This simulation assumes a $\Lambda$CDM cosmology with parameters based on the WMAP1 best-fit cosmology \citep{Spergel2003}:~$\sigma_8=0.9$, $n_s=1$,  $\Omega_{\rm m} = 0.25$, $\Omega_{\rm b}=0.045$, $\Omega_{\Lambda}=0.75$, and $h=0.73$ (where $H_{0} = 100$ $h$ km s$^{-1}$ Mpc$^{-1}$).  Millennium has a comoving box size of $500\,h^{-1}$\,Mpc with $2160^3$ particles of mass $8.6 \times 10^8\, h^{-1}\, {\rm m}_{\odot}$.  The minimum mass for a (sub)halo in the merger trees is 20 times the particle mass.

\ds~belongs to the SAGE (Semi-Analytic Galaxy Evolution) family of models, building on \citet{Croton2016}, but with the addition of many novel features.  Where classical semi-analytic models describe the stellar and gas discs of galaxies each with a singular reservoir with an analytic density profile, discs in \ds~are instead split into 30 annuli, allowing the density profile of the galaxy to be built numerically.  The model accounts for what is understood to be the core astrophysical processes relevant to galaxy evolution:~gas cooling and accretion, star formation and stellar feedback, central black-hole growth and associated feedback from active galactic nuclei, gravitational disc instabilities, galaxy mergers, and environmental stripping of gas.  Within each disc annulus, gas is broken into its ionized, atomic, and molecular phases \citep[based on][]{McKee2010,Fu2013}.  Star formation, stellar feedback, and disc instabilities all take place locally, affecting the structure of the gas disc and therefore its overall \HI~content (of central relevance to this work).

The 8 free parameters of this version of \ds~were manually calibrated to reproduce a set of observables.  Most notably for this work, these include the \HI~mass function \citep[based on data from][]{Zwaan2005,Martin2010}, the stellar mass function \citep{Baldry2008}, and the mean \HI-to-stellar mass ratio of galaxies as a function of stellar mass \citep{Brown2015} (all three of these are at $z\!\simeq\!0$).  Further details can be found in appendix B of \citet{Stevens2018}.

To minimise our results from being biased by unresolved systems in our sample, we exclude galaxies from our \ds~catalogue that reside in (sub)halos with a maximum historical mass%
\footnote{The `maximum historical mass' is the greatest mass of the (sub)halo found from walking back in time along the primary-progenitor branch of that subhalo in its merger tree of the Millennium simulation.}
less than the equivalent of 100 particles ($8.6 \times 10^{10}\, h^{-1}\, {\rm m}_\odot$), in line with \citet{Onions2012}. These systems are excluded in all parts of this paper, including where we present (sub)halo-specific results in Section \ref{sec:Xigg}.  Given this halo mass limit, we argue that galaxies with an \HI~mass above $10^9\,h^{-1}{\rm m}_\odot$ are those that are sufficiently complete in terms of halo mass and therefore trustworthy for the purposes of this work. We expect galaxies above these masses to be the most relevant for clustering (as they can be detected to greater distances, and hence increase the cosmological volume and clustering signal-to-noise) and Tully--Fisher measurements (as they are the highest signal-to-noise objects with the best measured rotation measurements). For example, this limit is below the mean and peak \HI~masses of objects detected in the WALLABY pilot surveys (Westmeier et.~al., in prep.; Courtois et.~al., in prep.).
 

 Although the version of {\sc Dark Sage} we are using here is built on a relatively old N-body simulation (Millennium), we use this as it is well tested and characterised in the literature \citep{Stevens2018,Stevens2019b}, which makes for easier understanding of the analytic models we develop here. As the aim of this work is to set up the first stages for a flexible \HI-based HOD, we do not think it is particularly important which simulation is used here, and defer a detailed look at the impact of different simulations and simulation choices for later work.


\section{The H\thinspace{\protect\scriptsize I} mass conditional HOD} \label{sec:hod}

\subsection{The standard Halo Occupation Distribution model} 

The standard HOD model contains the following three aspects: 
1) the number of galaxies in a parent halo; 
2) the spatial distribution of the galaxies in their parent halo;
3) the velocities of these galaxies. 
The last two aspects are modeled by the halo density profile, for example, the Navarro–Frenk–White (NFW) profile \citep{NFW1996,Navarro1997}. While the first aspect, i.e.~predicting how many galaxies are in a parent halo is the main goal of the study of HOD, and therefore is what the term HOD commonly refers to. 

Let $f_h(M) dM$ denote the number of parent halos in an infinitesimally small halo mass bin $[M,M+dM]$, where $M$ denotes (log) halo mass. $f_h(M)$ is the halo mass function, which has been well characterised from simulations (i.e., \citealt{Jenkins2001, Tinker2005}), 
but which for now we leave unspecified. If there are $dn_{c}$ central galaxies in these $f_h(M) dM$ parent halos, and $dn_{s}$ satellite galaxies, the HOD of central and satellite galaxies of that halo mass bin are defined as
\begin{subequations}
 \be \label{eq:hoddef}
\langle N_{\mathrm{cen}}(M)  \rangle \equiv \frac{dn_{c}}{f_h(M)\, dM}~,
\ee
\be \label{eq:hoddef_sat}
\langle N_{\mathrm{sat}}(M)  \rangle \equiv \frac{dn_{s}}{f_h(M)\, dM}~,
\ee 
\end{subequations}
respectively. The physical meaning of the HOD is that if the mass of a parent halo is $M$, the probability that it hosts a central galaxy is $\langle N_{\mathrm{cen}}(M)  \rangle$ of Eq.~\ref{eq:hoddef}, while the probability that it hosts a satellite galaxy is $\langle N_{\mathrm{sat}}(M)  \rangle$ of Eq.~\ref{eq:hoddef_sat}. A parent halo can host only one central galaxy at most, but multiple satellite galaxies, typically increasing in number with halo mass.
 
There are several commonly used HOD models in past works. 
\citet{Berlind2002} model $\langle N_{\mathrm{sat}}\rangle$ using a simple power-law function and assuming $\langle N_{\mathrm{cen}}  \rangle=1$, i.e.~each parent halo hosts one central galaxy.
\citet{Tinker2005} models $\langle N_{\mathrm{sat}}\rangle$ using a exponential function multiplied by a power law  function, while retaining $\langle N_{\mathrm{cen}}\rangle=1$.
\citet{Zheng2007b} models $\langle N_{\mathrm{cen}}\rangle$ using the error function, while modeling $\langle N_{\mathrm{sat}}\rangle$ using a power law function multiplied by the error function.  This model has been commonly used in previous research \citep{Guo2014,Guo2015,Manera2015,Howlett2015} to generate mock galaxy surveys. 

These commonly used HOD models have been shown to work poorly for H\thinspace{\protect\scriptsize I} \citep{Calette2021,Chauhan2021,Hadzhiyska2021} or more complex types of galaxies, such as Emission Line Galaxies \citep{GonzalezPerez2018,Avila2020}. Therefore, in this paper, we develop a new HOD model that adds another dimension to the standard HOD, based on the predictions of the {\sc Dark Sage} simulation.

\subsection{The definition of the  H\thinspace{\protect\scriptsize I} mass conditional HOD}

We can expand our definitions of $\langle N_{\mathrm{cen}} \rangle$ and $\langle N_{\mathrm{sat}} \rangle$ to represent the probability of a halo to host a central or satellite galaxy of a given H\thinspace{\protect\scriptsize I} mass using
\begin{subequations}
\label{defHODss}
\be 
\langle N_{\mathrm{cen}}(M,M_{\rm HI}) \rangle \equiv \frac{dn_{c,\rm HI}}{f_h(M)\, \varphi(M_{\rm HI})\, dM\, dM_{\rm HI}}~,
\ee
\be
\langle N_{\mathrm{sat}}(M,M_{\rm HI}) \rangle \equiv \frac{dn_{s,\rm HI}}{f_h(M)\, \varphi(M_{\rm HI})\, dM\, dM_{\rm HI}}~,
\ee 
\end{subequations}
where $dn_{c,\rm HI}$ and $dn_{s,\rm HI}$ denote the number of galaxies in an infinitesimally small 2-dimensional bin bounded by $[M,M+dM]$ and $[M_{\rm HI},M_{\rm HI}+dM_{\rm HI}]$. $\varphi(M_{\rm HI})$ denotes the H\thinspace{\protect\scriptsize I} mass function \citep{Zwaan2003, Zwaan2005, Jones2018}.

Defined in this way, the H\thinspace{\protect\scriptsize I}-conditional HOD reduces down to the standard HOD defined in Eq.~\ref{eq:hoddef} through integration over H\thinspace{\protect\scriptsize I} mass. Similar integration over halo mass returns the H\thinspace{\protect\scriptsize I} mass function, and integrating over both simply returns the total number of galaxies in the sample.

\begin{figure} 
\centering
 \includegraphics[width=\columnwidth]{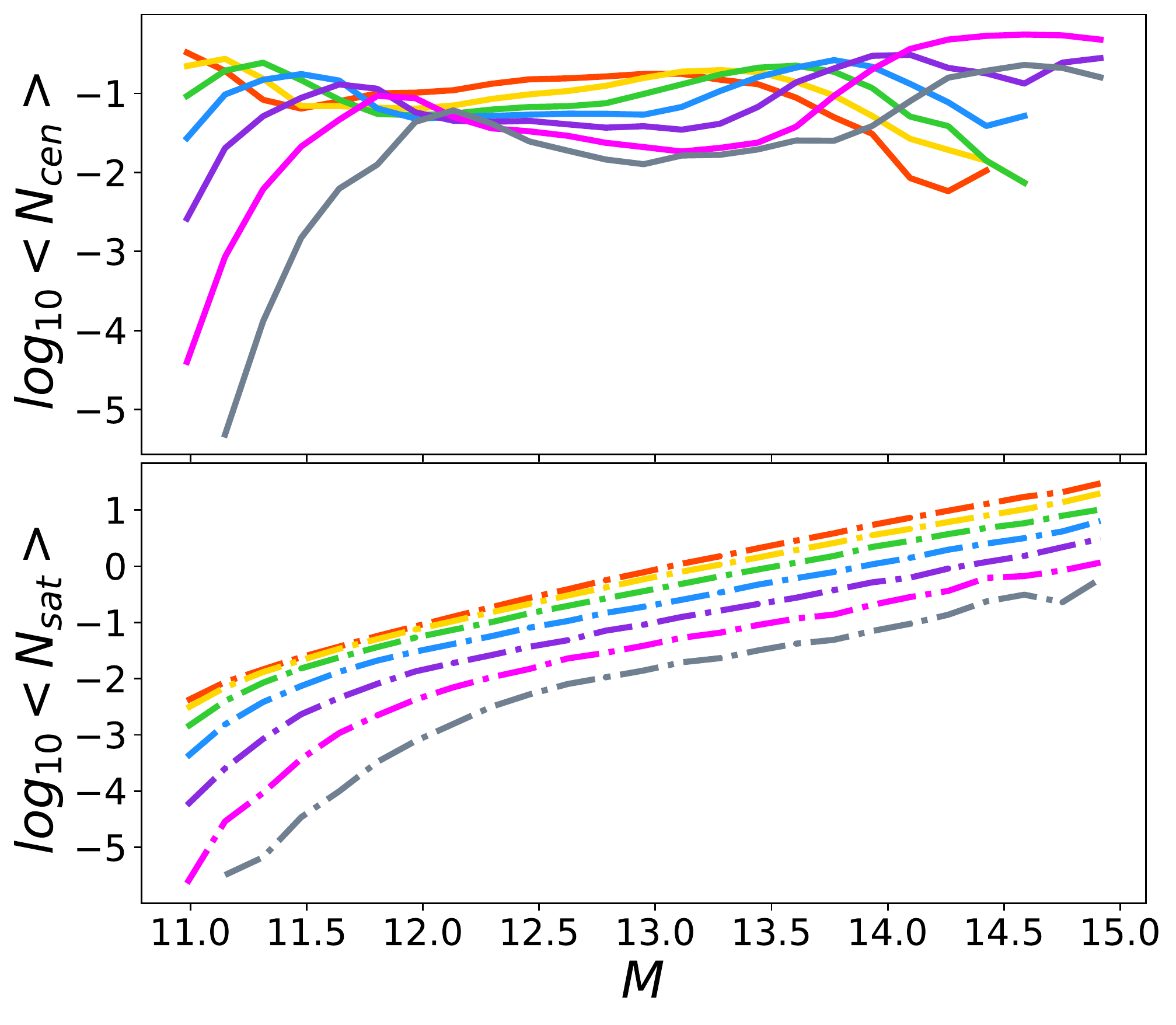}
 \includegraphics[width=\columnwidth]{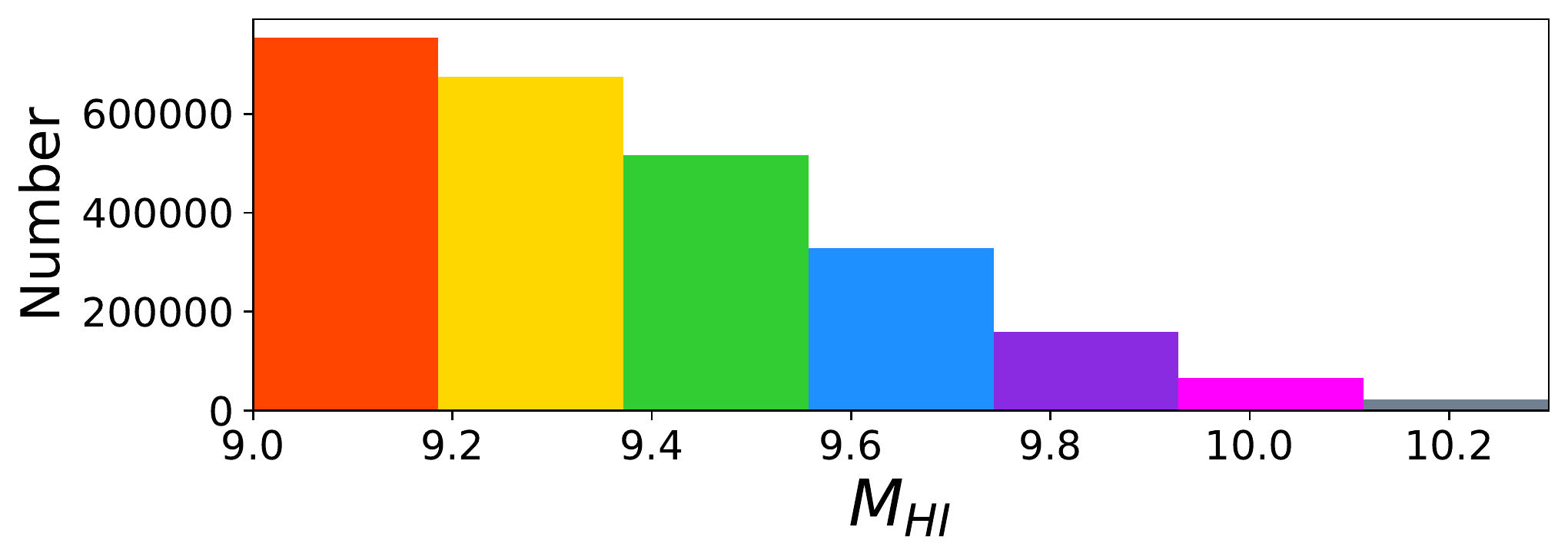}
 \caption{ The HOD of H\thinspace{\protect\scriptsize I} galaxies { as measured from the output of {\sc Dark Sage}}. The top panel showcases the HOD of central galaxies: 
 { that is, the probability that a halo of a given mass (per the $x$-axis) will host a central galaxy of a given \HI~mass.}
 The middle panel showcases the HOD of satellite galaxies:
 {  i.e., the expected number of satellites within a given \HI~mass bin to exist in a halo of a given mass.}
 The color of the curves {in the   top and middle panels} indicate the H\thinspace{\protect\scriptsize I} mass of the galaxies, { corresponding to the bar of the same color in the bottom panel}.}
 \label{HIHOD}
\end{figure}

The top two panels of Fig.~\ref{HIHOD} display the HOD of H\thinspace{\protect\scriptsize I} galaxies measured from the {\sc Dark Sage} simulation. The top panel shows the HOD of the central galaxies, the middle panel shows the HOD of the satellite galaxies. The colors of the curves indicate the H\thinspace{\protect\scriptsize I} mass of the galaxies, based on the histogram of the bottom panel.   In this sense, Fig.~\ref{HIHOD} is effectively a normalised 2-D histogram across H\thinspace{\protect\scriptsize I} and halo mass (split additionally into central and satellite populations) measured directly from {\sc Dark Sage}.

The histogram in the bottom panel of Fig.~\ref{HIHOD} only displays the distribution of H\thinspace{\protect\scriptsize I} mass above $10^{9}h^{-1} m_{\sun}$, which is the  effective resolution limit of our {\sc Dark Sage} catalogue. As a result of this resolution limit, the sum of $\langle N_{\mathrm{cen}}(M,M_{\rm HI}) \rangle$ over all H\thinspace{\protect\scriptsize I} mass bins we consider is less than 1, even though every halo used in {\sc Dark Sage} is assigned a central galaxy. We confirmed that including H\thinspace{\protect\scriptsize I} mass bins lower than $10^{9} h^{-1} m_{\sun} $ recovers a total number of central galaxies equal to the number of halos, but do not enforce this condition or use such low H\thinspace{\protect\scriptsize I} mass bins in the remainder of our analysis as their clustering and distribution is less reliable.

\subsection{The H\thinspace{\protect\scriptsize I} mass conditional HOD model}

From the top panel of Fig.~\ref{HIHOD}, we find that the HOD measured from the central galaxies showcases a characteristic `M' shape as a function of halo mass, but with a position and size of `dip' that depends on the H\thinspace{\protect\scriptsize I} mass we consider. In intermediate-mass halos, i.e. around $11.8<M<13.5$, this is caused by AGN feedback `turning on' which slows down the gas cooling and hence the replenishment of H\thinspace{\protect\scriptsize I} gas within central galaxies (see Section 4.1.1 of \citealt{Chauhan2020} for more discussion). Therefore the HOD of central galaxies is mildly decreased in this region. 
In low-mass halos the number of high H\thinspace{\protect\scriptsize I} mass centrals decreases rapidly as the H\thinspace{\protect\scriptsize I} mass approaches a significant fraction of the halo mass, due to the inability of these halos to retain and cool the gas.

The HOD for satellites follows a more usual power-law shape, with high-mass halos hosting more satellites, but again with an exponential decline towards lower halo masses due to the relative inability of galaxies in these halos to retain, accrete and cool new material. 
The amplitude of the satellite HOD increases with decreasing H\thinspace{\protect\scriptsize I} mass --- to be expected given the shape of the H\thinspace{\protect\scriptsize I} mass function and relative abundances of low vs. high H\thinspace{\protect\scriptsize I} mass galaxies --- however 
there is also evidence that higher-H\thinspace{\protect\scriptsize I}-mass satellites experience a stronger exponential decrease in their abundance as we go to lower halo masses.

To model the HOD of the central and satellite galaxies, we take inspiration from \citet{Weigel2016}, using double and single Schechter functions for the central and satellite HODs, respectively:
\begin{subequations}
\be \label{HODcen6}
\begin{split}
\langle N_{\mathrm{cen}}(M,M_{\rm HI})\rangle = & \phi_c 10^{\beta_1(M_c-M)}\exp{\left[ -10^{(M_c-M)}\right]}\\
+ & \phi_a 10^{\beta_2(M-M_a)}\exp{\left[ -10^{(M-M_a)}\right]}~,
\end{split}
\ee 
\be \label{HODsat}
\langle N_{\mathrm{sat}}(M,M_{\rm HI}) \rangle = \phi_s 10^{\beta_s(M_s-M)}\exp{\left[ -10^{(M_s-M)}\right]}~,
\ee
\end{subequations}
where the parameters $M_c$,  $\beta_1$, $\beta_2$, $\phi_c$, $\phi_1$, $M_a$ and $M_s$, $\beta_s$, $\phi_s$ are functions of H\thinspace{\protect\scriptsize I} mass. However, as we show in Section~\ref{sec:resultsimplify} there can be strong correlations between these parameters that allow us to reduce the model dimensionality. They can be fitted from the measured HOD or the galaxy two-point correlation function. How this can be done is the subject of the next section.

\section{The correlation function and the halo distribution} \label{sec:Xigg}

In this section we show how the  HOD can be fit by comparing a model galaxy two-point correlation function to the measured correlation function of a real galaxy survey.%
\footnote{In this paper, the `real galaxy survey' has been replaced with a simulated data set from a semi-analytic model, but we expect the methodology to remain valid.} 
To achieve this goal, we need to firstly write down the model galaxy two-point correlation function in terms of the HOD, then compare the model to the measurement to determine the HOD parameters.      
 
The galaxy two-point correlation function can be decomposed into the sum of a `one-halo' term, $\xi^{1h}_{gg}$, and a `two-halo' term, $\xi^{2h}_{gg}$, written as \citep{Berlind2002,Yang2003,Zheng2004}
\be 
\xi_{gg}(r)=[1+\xi^{1h}_{gg}(r)]+\xi^{2h}_{gg}(r)~,
\ee 
where $r$ is the pair separation.
The one-halo term is the contribution from intrahalo pairs: i.e., two satellites, or one central and one satellite, situated within the same halo.
The two-halo term is the contribution from interhalo pairs: i.e., two galaxies, centrals or satellites, in their own unique halos.

\subsection{The One-Halo Term}\label{1halotm}
The model of the one-halo term can be expressed as  \citep{Berlind2002,Tinker2005,Zheng2016}
\be \label{eq:haloone}
\begin{split}
 1+\xi^{1h}_{gg}(r)=&\frac{1}{  2\uppi\, r^2\, \bar{n}_g^2 }\int_0^{\infty} \frac{1}{\alpha R_{\mathrm{vir}}}
  \bigg [
 \left\langle N_{\mathrm{cen}}\, N_{\mathrm{sat}}\right\rangle\, f_{cs}(x) \\
 +&
 \frac{\left\langle N_{\mathrm{sat}}(N_{\mathrm{sat}}-1)\right\rangle }{2} f_{ss}(x)
 \bigg ] f_{h}(m)\, dm ~,
 \end{split}
\ee 
where $x\equiv\frac{r}{\alpha R_{\mathrm{vir}}}$, $f_{h}(m)$ is the halo mass function, $\bar{n}_g$ is the average number density of galaxies. $\alpha$ is a parameter chosen to set the maximum extent of what we consider to be within `one halo'. We also assume that the number of galaxies follows the Poisson distribution, $\langle N_{\mathrm{sat}}(N_{\mathrm{sat}}-1)\rangle =\langle N_{\mathrm{sat}} \rangle^2$ \citep{Berlind2002,Tinker2005}. $\bar{n}_g$ 
can be calculated using \citep{Yang2003,Tinker2005}
\be 
\bar{n}_g=\int_0^{\infty} \left[\langle N_{\mathrm{cen}}\rangle + \langle N_{\mathrm{sat}}\rangle\right]\,  f_{h}(m)\, dm~.
\ee
For the {\sc Dark Sage} simulation, $\bar{n}_g$ can also be simply calculated using $\bar{n}_g = N_g/V$, where $N_g$ is the total number of galaxies in the simulation and $V$ is the volume of the simulation box.

 The virial radius of a halo, $R_{\mathrm{vir}}$ is computed from the halo mass using 
\be \label{Rvirs}
R_{\mathrm{vir}}=\left( \frac{3 m}{4\uppi\, \Delta\, \rho_c}\right)^{1/3} ~,
\ee 
where $\Delta=200$ is the overdensity threshold of halo identification,   consistent with the definition of halo (virial) mass in the simulation. 
$\rho_c=
 3H_0^2/(8\uppi G)$ is the  critical density of the (simulated) universe, where $G$ is Newton's gravitational constant and $H_{0}$ the Hubble constant.

$f_{cs}(x)$ expresses the radial distribution of central--satellite galaxy pairs in a parent halo \citep{Zheng2004,Tinker2005,Zheng2007}, which, as the central galaxy is presumed to sit in the center of mass of the halo, is usually written in terms of halo radial density profile $\rho_m(r)$
\be\label{fcseq}
f_{cs}(x)\propto \rho_m(r)\, r^2~. 
\ee 
$f_{ss}(x)$ is the probability density function (PDF) of satellite galaxy pairs in a parent halo. This is more difficult to express, and requires convolving a given halo density profile with itself. It can be done analytically for the truncated NFW density profile (and others, see appendix A of \citealt{Zheng2007}) although the resulting expressions are cumbersome and so not repeated here.
As PDFs, both $f_{cs}(x)$ and $f_{ss}(x)$ should be normalized to 1 when integrating from $x=0$ to $x=1$
\citep{Zheng2004,Tinker2005,Zheng2007,Zheng2016}.

It is important to note that Eq.~\ref{eq:haloone} is general and can apply for any radial distribution or (radially symmetric) halo profile. However, the commonly used forms for $f_{cs}(x)$ and $f_{ss}(x)$ are for the distribution of dark matter within halos. As we are interested in the \textit{galaxy} correlation function, these may not be appropriate to use if the distribution of galaxies differs significantly from the dark matter. We investigate this for our simulated H\thinspace{\protect\scriptsize I} galaxies in Section~\ref{sec:app}.

\subsection{The Two-Halo Term}

The two-halo term $\xi^{2h}_{gg}(r)$ is the contribution from interhalo galaxy pairs. The basic assumption behind this term is that the clustering of galaxies can be written in terms of the underlying matter clustering multiplied by the galaxy bias, $b_{g}(k)$, which here is written as a function of the Fourier scale $k$. The galaxy bias then stems from the underlying halo bias function $b_{h}(m)$, weighted by the occupancy of each halo, and the (Fourier-space) halo density profile \citep{Berlind2002}.

However, following the arguments in \citet{Zheng2004}, at any given separation it is important to only include galaxy pairs arising from two different distinct halos. At separations within the transition region from the one- to two-halo terms it becomes important to consider that \textit{two} halos simply may not exist --- their extent is such that they would have merged to form a single larger halo. This concept, called `halo exclusion', can be incorporated into the model by requiring that the pair separation $r$ must be larger than double the halo's virial radius, i.e.~$r>2R_{\mathrm{vir}}$. 
Using Eq.~\ref{Rvirs}, one can then find the following mass limit   
\citep{Zheng2004, Tinker2005}
\be 
m_{\mathrm{lim}}(r)=  \frac{4}{3}\uppi\left( \frac{r}{2} \right)^3 \rho_c\, \Delta  ~,
\ee
which can be used to enforce that only if the halo mass $m$ is less than $m_{\mathrm{lim}}(r)$, should it be included into the two-halo term for pair separation $r$.

From these considerations, the model of the two-halo term $\xi^{2h}_{gg}(r)$ is given by \citep{Zheng2004,Tinker2005,Zheng2007}
\be 
\xi^{2h}_{gg}(r)= \left[1+\xi'_{2h}(r)\right]\left( \frac{\bar{n}'_g}{\bar{n}_g} \right)^2 -1 ~,
\ee 
where, 
\be 
\bar{n}'_g = \int_0^{m_{\mathrm{lim}}(r)} \left[\langle N_{\mathrm{cen}}\rangle + \langle N_{\mathrm{sat}}\rangle\right]\,  f_{h}(m)\, dm ~,
\ee 
is the average number density of galaxies that reside in halos with $m\leq m_{\mathrm{lim}}$. The un-normalised correlation function, $\xi'_{2h}(r)$, is computed by inverse Fourier transforming the galaxy power spectrum, such that
\be \label{xigg2h}
\xi'_{2h}(r)=\frac{1}{2\uppi^2}\int_0^{\infty} b_{g}^{2}(k,r)\, P_m(k)\, k^2\, \frac{\sin(kr)}{kr}\, dk ~,
\ee 
where $P_m(k)$ is the nonlinear matter power spectrum (which we generate using the \verb~CAMB~ package;  \citealt{Lewis:1999bs,Howlett2012}). 
The galaxy biasing parameter $b_g$ is given by
\be \label{bgm}
b_g=\frac{1}{\bar{n}'_g}\int^{m_{\mathrm{lim}}(r)}_0 \left[ \langle N_{\mathrm{cen}} \rangle + \langle N_{\mathrm{sat}}  \rangle   \right] b_h(m)\, y_g(k)\, f_{h}(m)\, dm ~,
\ee 
where $b_h(m)$ is the halo biasing parameter, see Appendix \ref{sec:bh}. $y_g(k)$ is the Fourier transformation of $\rho_m$ of Eq.~\ref{fcseq}, defined as \citep{Cooray2002}
\be \label{ygeq}
y_g(k)\equiv\int_0^{R_{\mathrm{vir}}} 4\uppi R^2 \frac{\sin(kR)}{kR}\frac{\rho_m(R)}{m}dR~.
\ee
Assuming the galaxies follow the NFW profile, $y_g$ is given by Eq.~\ref{ygnfw}.  

The model described above is perhaps the simplest model one can consider that incorporates halo-exclusion. However, it does not account for the cross-correlation between two halos where only one has mass less then $m_{\mathrm{lim}}$. It also does not account for the ellipsoidal nature of halos. \cite{Tinker2005} discuss both of these issues and explore more sophisticated treatments that can account for them. However, for this first work we consider the simpler and less computationally intensive model.

In the next section, we use our model for the correlation function to fit for the HOD parameters in the \textsc{Dark Sage} simulation. We also explore some of the assumptions in the modelling, particularly the intrahalo distributions of H\thinspace{\protect\scriptsize I} galaxies, and characterise how well they represent our simulation.

\section{Application to the full {\sc Dark Sage} Simulation}\label{sec:app}

\subsection{Validating the radial profiles}
\label{sec:radialprofile}

We begin by measuring the basic ingredients going into our model of the correlation function, the $f_{cs}$, $f_{ss}$ and $y_g$ profiles of the H\thinspace{\protect\scriptsize I} galaxies in the {\sc Dark Sage} simulation, then compare them to the NFW halo density profile.

\begin{figure} 
\centering
 \includegraphics[width=\columnwidth]{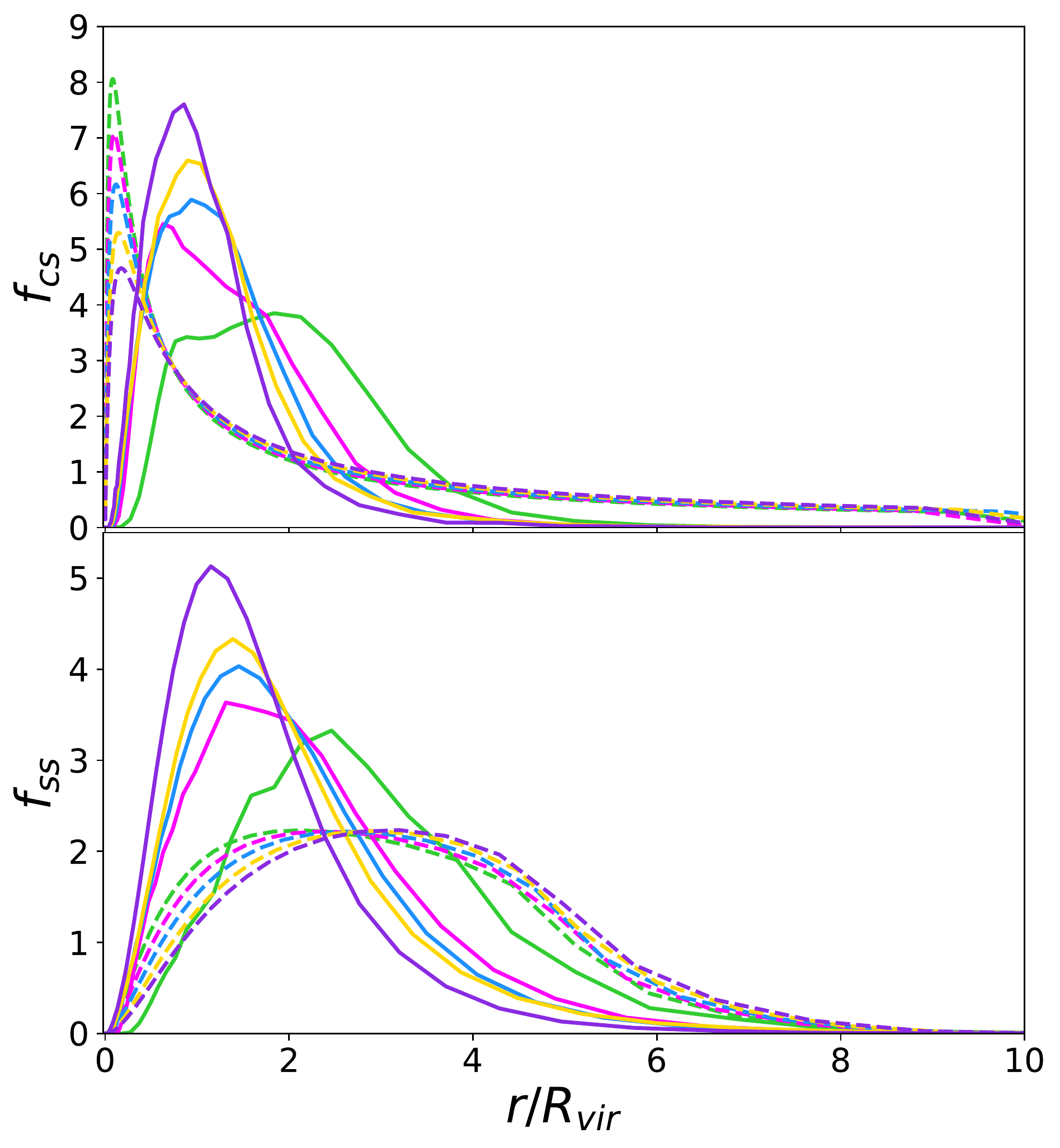}
 \caption{The $f_{cs}$ and $f_{ss}$ for five example halo mass bins.  The green, pink, blue, yellow and purple colored lines are for halo mass bins centered on $M=11.31$, $12.13$, $12.95$, $13.77$ and $14.59$ respectively. The dashed curves depict the NFW profile for these masses. The solid curves showcase the measurements from {\sc Dark Sage}. All these curves (both solid and dashed) are normalized to integrate to one in the interval  $x\in[0,1]$. } 
 \label{fcsfss}
\end{figure}

$f_{cs}$ and $f_{ss}$ are computed by splitting the halos into five different mass bins in the interval $M\in[10.9, 15]$ and counting the number of central--satellite and satellite--satellite pairs in these halos. We plot them against $r/R_{\rm vir}$ in Fig.~\ref{fcsfss}, normalized to integrate to one in the interval $x\in[0,1]$ with $\alpha=10$. 
The $f_{cs}$ and $f_{ss}$ curves derived from the NFW profile are  also shown in the figure with the same normalisation. As can be seen, there is poor agreement between the measurements of the pairwise distributions of the H\thinspace{\protect\scriptsize I} galaxies (which directly traces the distribution of resolved subhalos in the Millennium simulation) from {\sc Dark Sage} and an NFW profile, even if we allow for galaxies separated as far apart as $10\,R_{\rm vir}$ (where, typically, one would invoke an upper limit of $2\,R_{\rm vir}$).

Clearly, Millennium subhalos (and therefore {\sc Dark Sage} satellite galaxies) are mostly found at separations from their central around or beyond the virial radius of the halo. A ``halo'' from the perspective of the framework we are using here is just a friends-of-friends group of particles \citep{Springel2005}. This has a calculable $R_{\rm vir}$, but the group itself can extend well beyond the virial radius. The peak distance for subhalos is around $R_{\rm vir}$ because close to the halo center, subhalos can merge with the central, get disrupted, or become harder to identify, which can result in their removal from the merger trees used as an input for a semi-analytic model \citep[see e.g.][]{Onions2012, Poulton2020}. 

It is also important to note that part of the reason why there are fewer satellite galaxies closer to halo centers in {\sc Dark Sage} is also because there are no ``orphan'' galaxies (that is, galaxies that have lost their dark-matter subhalo). For an H\thinspace{\protect\scriptsize I} study, the lack of orphans is less of a cause for concern than it would be for an optical study, because if a subhalo is tidally stripped to the point that it is no longer resolvable in the $N$-body simulation, then the gas in that subhalo's galaxy would also be quite susceptible to stripping processes too \citep[cf.~the clustering results of][]{Knebe2018}. Hence, we are making the implicit, but physically reasonable, assumption in this study that galaxies without appreciable dark matter also lack appreciable gas through tidal forces and ram-pressure stripping.

A secondary trend can also be seen in Fig.~\ref{fcsfss}, where lower-mass halos typically host satellites at larger distances. This is because the merger time-scales are shorter for low-mass halos, where the masses of the merging systems should statistically be more comparable.  High-mass halos can host satellites stably orbiting at a range of distances because the satellites are often relatively small, to the point that they have next to no gravitational influence on the rest of the halo.

As a result, it is clear that enforcing all H\thinspace{\protect\scriptsize I} galaxies to be within $2R_{\rm vir}$ and to follow $f_{cs}$ and $f_{ss}$ derived from an NFW profile is inappropriate. We find that to cover all the measured $f_{cs}$ and $f_{ss}$ curves, the normalization interval should allow for separations up to $10R_{\rm vir}$. To do this, we set $\alpha=10$ in our model for the one-halo term of the correlation function Eq.~\ref{eq:haloone} going forward, and normalize the measured $f_{cs}$ and $f_{ss}$ of {\sc Dark Sage} simulation in $x\in[0,1]$. We then use the measured curves as input for our modelling.

\begin{figure} 
\centering
 \includegraphics[width=\columnwidth]{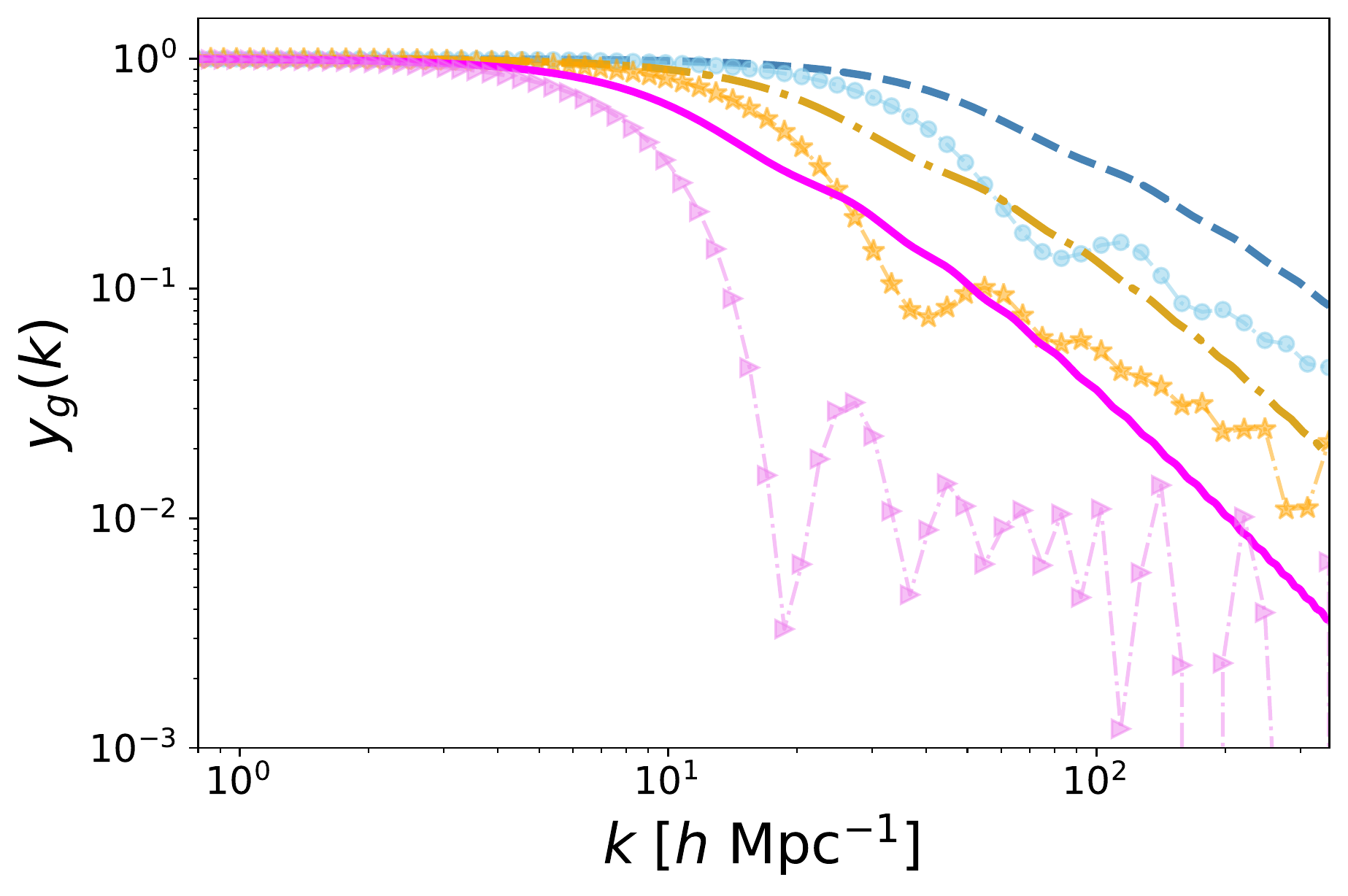}
 \caption{The dashed, dash-dotted and solid curves depict the Fourier transformation of the NFW profile calculated using Eq.~\ref{ygnfw}.  The dots, stars and triangular markers showcase the Fourier transformation of the measured galaxy radial distribution from the  {\sc Dark Sage} simulation. The blue color is for halo mass $M=11$, the brown color is for halo mass $M=12$, the pink color is for halo mass $M=13$.}
 \label{ygk}
\end{figure}

We also measure $y_g$ of the H\thinspace{\protect\scriptsize I} galaxies in the {\sc Dark Sage} simulation using the same halo mass bins as above.
We do this firstly by calculating $\rho_{g}$, the galaxy density profile, from the measured $f_{cs}$ using Eq.~\ref{fcseq}.  
Then we calculate $y_g$ using Eq.~\ref{ygeq}, normalizing it to 1 at $k=0\,h\,\mathrm{Mpc}^{-1}$. These measurements are shown in Fig.~\ref{ygk}, alongside the $y_g(k)$ predicted from the NFW profile (calculated using Eq.~\ref{ygnfw}) for our three example halo mass values. Again, we find that the NFW profile prediction is different from the $y_g$ measured from the {\sc Dark Sage} simulation, with less `power' on small scales (large $k$). This is indicative of greater distances between H\thinspace{\protect\scriptsize I} galaxies and the centers of their parent halos than would be predicted for dark matter. As such, we also use the measured values of $y_g$ in our correlation function modelling. 

Future work could aim to characterise and understand these trends in greater detail, and come up with flexible functional forms that could be incorporated into the HOD modelling as required. However, for the purposes of demonstrating that it is in principle possible to model the conditional HOD of H\thinspace{\protect\scriptsize I} galaxies using the clustering, we stick with the simulation-measured distributions here.

\subsection{Fitting an HOD for the whole {\sc Dark Sage} simulation}

In this section, we validate our fitting pipeline to show that we can fit the HOD of the whole {\sc Dark Sage} simulation \textit{without} splitting into H\thinspace{\protect\scriptsize I} mass bins. Fits to the HOD conditional on the H\thinspace{\protect\scriptsize I} mass $M_{\rm HI}$ are presented later in Section \ref{secHIHODss}.

 {\sc Dark Sage} requires that each parent halo hosts a central galaxy. Therefore, the HOD of central galaxies is given by $\langle N_{\mathrm{cen}} \rangle =1$.
We use the following equation to model the HOD of satellite galaxies  
\citep{Zheng2007b,Howlett2015,Zheng2016}
\be \label{satHODDS}
\langle N_{\mathrm{sat}}  \rangle = \langle N_{\mathrm{cen}}  \rangle \left( \frac{m-10^{M_{\mathrm{cut}}}}{10^{M_1}} \right)^{\beta} ~,
\ee 
where the parameters $M_{\mathrm{cut}}$, $\beta$ and $M_1$ are the HOD parameters that need to be fit by comparing the model correlation function $\xi_{gg}^{\rm mod}$ of Section \ref{sec:Xigg} to the measurement $\xi_{gg}^{\rm mea}$.  $M_{\mathrm{cut}}$ is the mass limit under which the halos are too small to host satellite galaxies. $M_1$ is the amplitude of the HOD of satellite galaxies. $\beta$ describes how fast the number of satellite galaxies increases as their parent halo mass increases; it is the slope of the HOD.  Due to the fact that the galaxies in {\sc Dark Sage} do not follow the NFW profile, we use the measured $f_{cs}$, $f_{ss}$, $y_g$ and $\alpha=10$ to calculate the model correlation function of Section \ref{sec:Xigg}. We split the halos in the interval $M\in[10.9, 15]$ with bin width $dM=0.036$. 

\begin{figure*} 
\centering
\includegraphics[width=81mm]{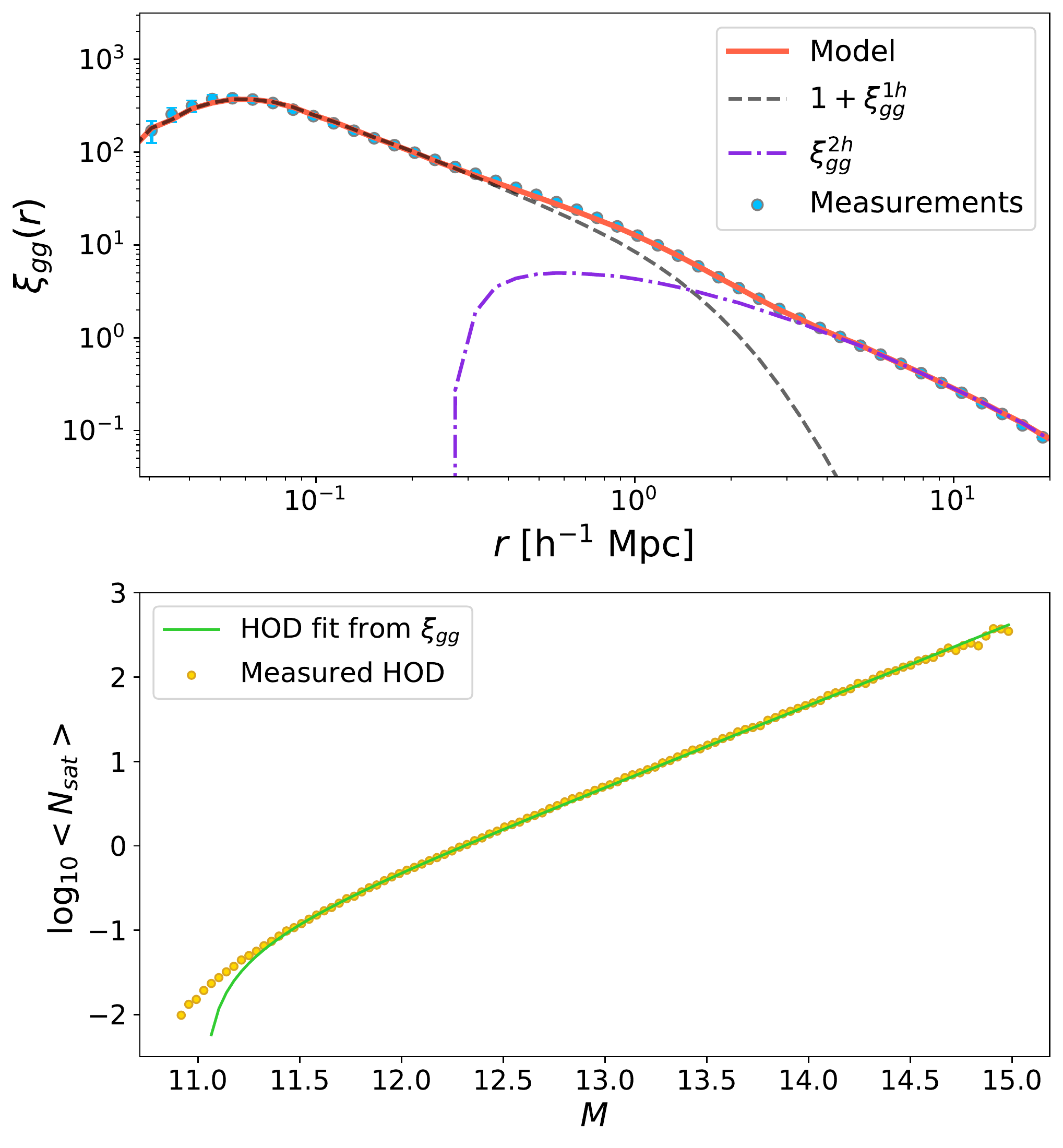}
 \includegraphics[width=89mm]{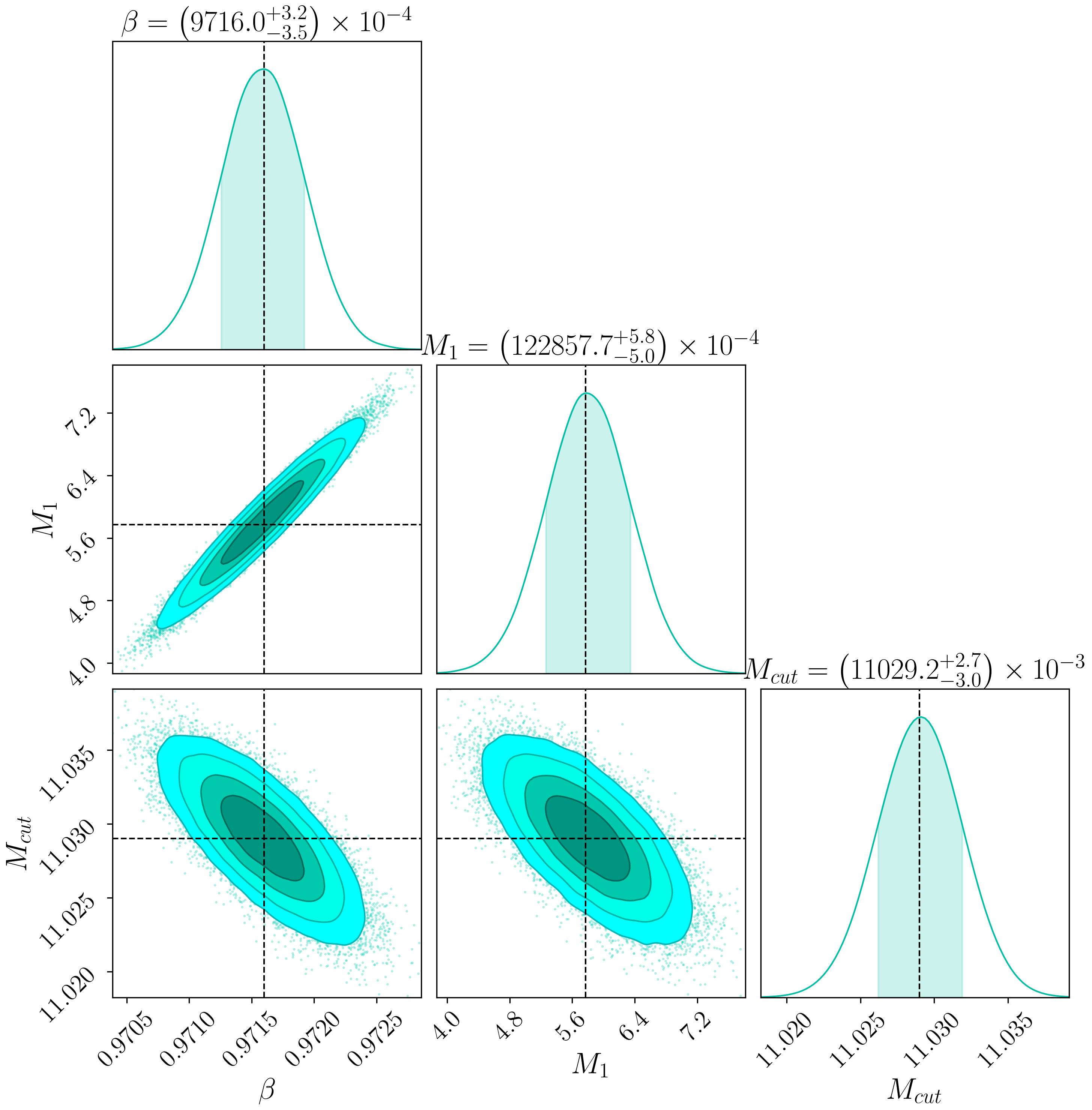}
 \caption{ Fitted HOD parameters from the galaxy two-point 
correlation function for the {\sc Dark Sage} simulation. In the left-hand panels, the top shows the correlation function measurement (blue-filled circles) and the model (red curve) fit to the measurement. This is further decomposed into the corresponding one-halo, $1+\xi_{gg}^{1h}$, and two-halo, $\xi_{gg}^{2h}$, terms (dashed-grey and dash-dotted purple respectively). The bottom panel shows the measured HOD and that inferred from the fitted correlation function (yellow points and green line respectively). The fit results of the HOD parameters are shown in the right-hand panels. The histograms show the distribution of the MCMC samples. The shaded areas in the 1D plots indicate the 68\% confidence level, while the 2D contours indicate the 1, 1.5, 2 and 2.5$\sigma$ regions.}
 \label{XiggHODs}
\end{figure*}

\begin{figure} 
\centering
 \includegraphics[width=\columnwidth]{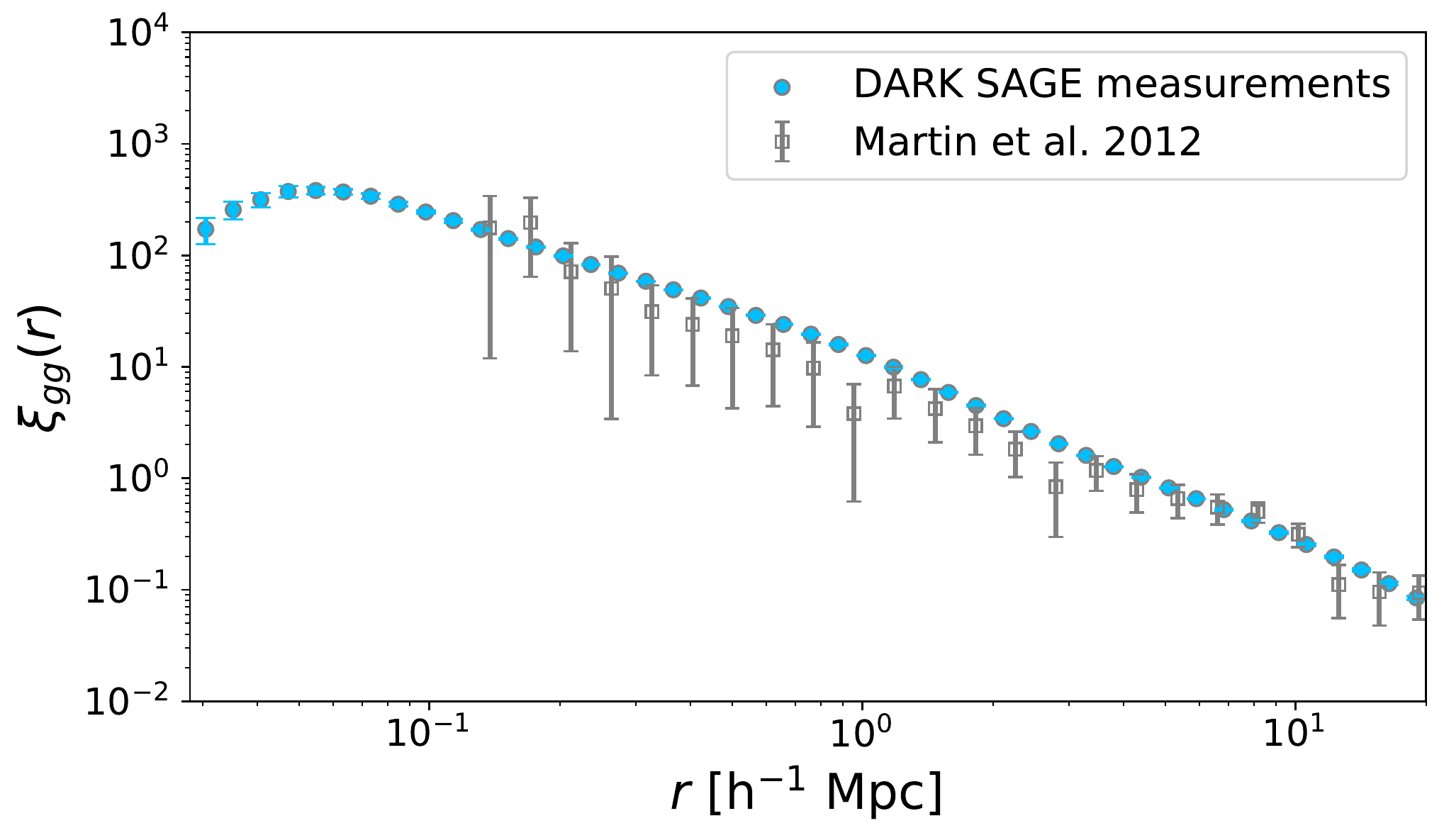}
 \caption{ { The comparison between the $\xi_{gg}^{\rm mea}$ of {\sc Dark Sage} (blue dots) and the observational measurements from \citet{Martin2012} (gray squares).} }
 \label{ygkcao}
\end{figure}

In the top-left panel of Fig.~\ref{XiggHODs}, the blue dots represent the measured galaxy two-point correlation function $\xi_{gg}^{\rm mea}$ of the {\sc Dark Sage} simulation.%
\footnote{We calculate this using the \verb~PYTHON~ package \verb~CorrFunc~ \url{https://corrfunc.readthedocs.io/en/master/}.} 
 Its comparison with the observational measurements from \citet{Martin2012} is shown in Fig.~\ref{ygkcao}. \citet{Martin2012} measures the \HI-selected galaxy two-point correlation function from the $\alpha.40$ sample of ALFALFA \citep{Haynes2011}. Our measurements are in agreement with \citet{Martin2012}, indicating that $\xi_{gg}$ from {\sc Dark Sage} serves as an appropriate proxy for that of an \HI~survey for our HOD-fitting method.

The HOD parameters are fitted by minimizing the
$\chi^2$ between $\xi_{gg}^{\rm mea}$ and $\xi_{gg}^{\rm mod}$, given by
\be\label{chi23ww}
\chi^2 = (\xi_{gg}^{\rm mea}-\xi_{gg}^{\rm mod})\boldsymbol{\mathsf{C}}^{-1}(\xi_{gg}^{\rm mea}-\xi_{gg}^{\rm mod})^T ,
\ee
where $\boldsymbol{\mathsf{C}}$ is the jackknife covariance matrix, expressed as \citep{Escoffier2016}
\be 
\boldsymbol{\mathsf{C}}_{ij}=\frac{N_s-N_d}{N_d\, N_{JK}}\sum^{N_{JK}}_{k=1} (\xi_{gg}^{{\rm mea},ik} - \bar{\xi}_{gg}^{{\rm mea},i})(\xi_{gg}^{{\rm mea},jk} - \bar{\xi}_{gg}^{{\rm mea},j})~.
\ee 
Following the arguments in \citet{Escoffier2016}, we divide the simulation box into $N_s=8$ identical sized cubes, then randomly delete $N_d=4$ cubes to obtain $N_{JK}=70$ jackknife samples. $\bar{\xi}_{gg}^{{\rm mea},i}$ is the average of $\xi_{gg}^{\rm mea}$ of the $N_{JK}$ jackknife samples in the $i^{\rm th}$ separation bin. 

The Metropolis--Hastings Markov chain 
Monte Carlo (MCMC) algorithm is applied to estimate the HOD parameters. We use flat priors in the interval $M_1\in[11,~13]$, $\beta\in[0.7,~1.5]$ and $M_{\mathrm{cut}}\in[10,11]$.

The fit results are shown in Fig.~\ref{XiggHODs}. The estimated HOD parameters are $\beta=0.9716^{+0.0032}_{-0.0035}$, $M_1   = 12.2858^{+0.0058}_{-0.0050}$ and $M_{\mathrm{cut}} = 11.0292^{+0.027}_{-0.030}$.
Plugging these best-fit values into Eq.~\ref{satHODDS}, we obtain the HOD of satellite galaxies predicted by the galaxy two-point correlation function, as shown in the bottom-left panel of Fig.~\ref{XiggHODs}. The predicted HOD from fitting the galaxy clustering is in excellent agreement with the measured HOD, which validates our pipeline and the use of the measured central and satellite profiles in our fitting.
In the bottom-left panel of Fig.~\ref{XiggHODs}, there is a slight disagreement between the measured and predicted HODs below $M=11.25$. This is due to the imperfect fit of the model correlation function in smaller separation bins ($r<0.05\, h^{-1}$\,Mpc), arising from the combination of measurement error in the correlation function \textit{and} in our theoretical model, which relies on noisy measurements of $f_{ss}$ and $f_{cs}$. The presence of noise in our theory prediction is not accounted for in the covariance matrix, and so leads to this small disagreement. However, this is a small enough effect that we do not deem it necessary to fix. If instead we were to use the analytic $f_{ss}$ and $f_{cs}$ of the NFW profile as the input of the model correlation function, the predicted HOD would have a significantly higher one-halo term, which would lead to a larger amplitude, $M_1$, and larger mass limit, $M_{cut}$, as well as smaller slope $\beta$. However, it more generally also leads to difficulty fitting both the one- and two-halo parts of the correlation function simultaneously.

\section{Application as a function of H\thinspace{\protect\scriptsize I} mass} \label{sec:Result}

In this section, we take our HOD fitting methodology, and apply it to different H\thinspace{\protect\scriptsize I} mass bins within the simulation.

\subsection{Checking the form of the H\thinspace{\protect\scriptsize I}-conditional HOD model} \label{sec:ResultHOD}

\begin{figure} 
\centering
\includegraphics[width=\columnwidth]{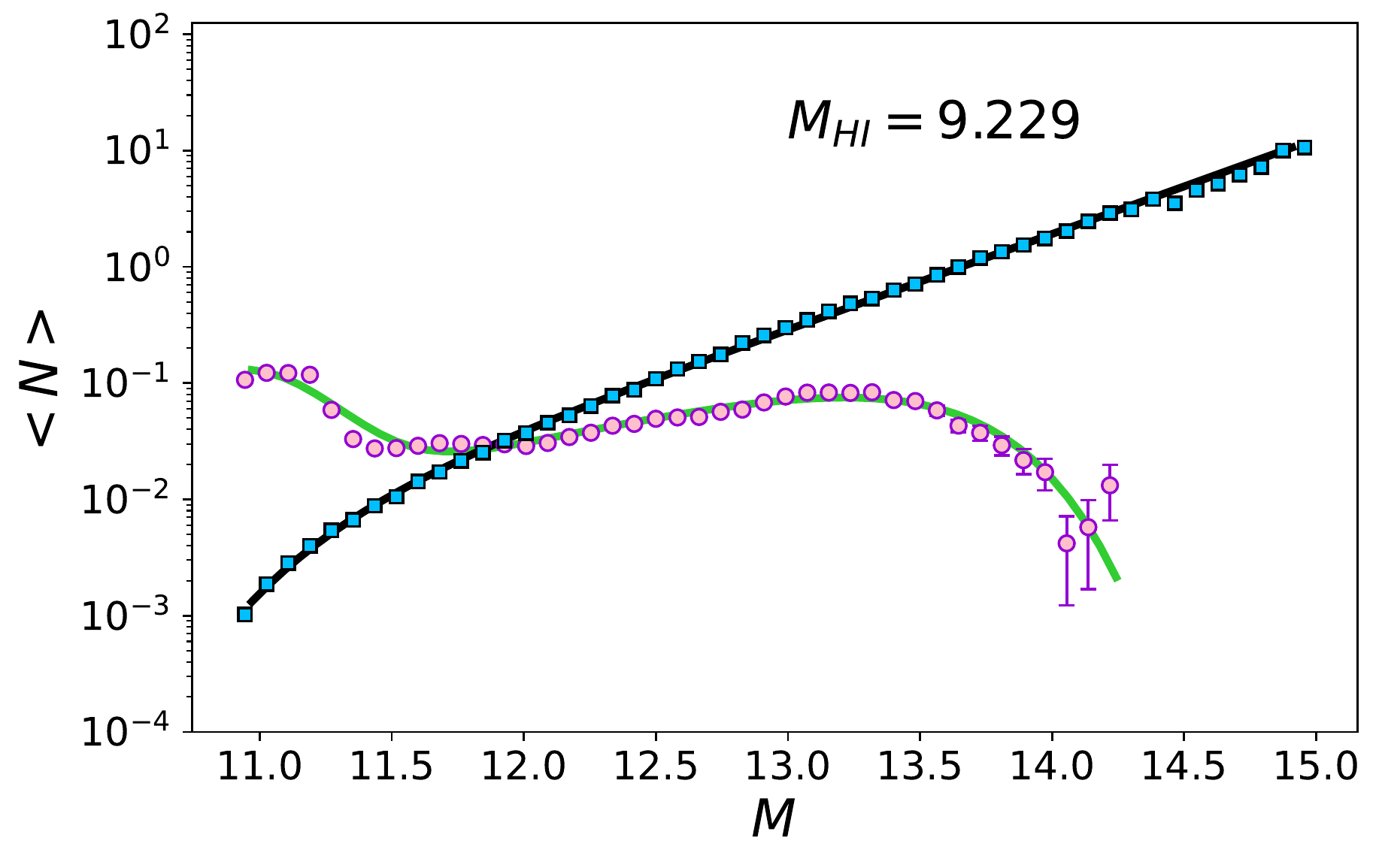}
 \includegraphics[width=\columnwidth]{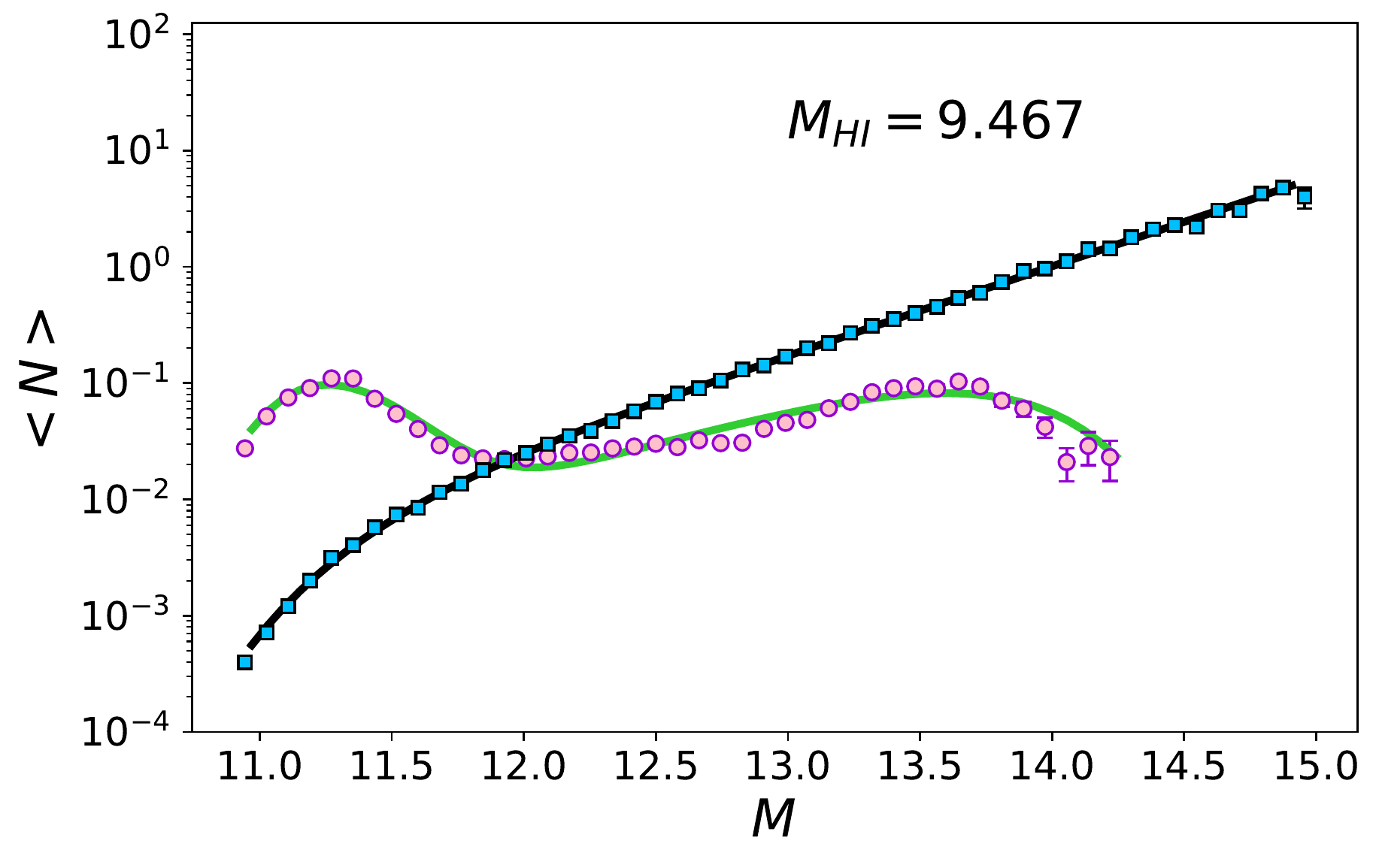}
 \includegraphics[width=\columnwidth]{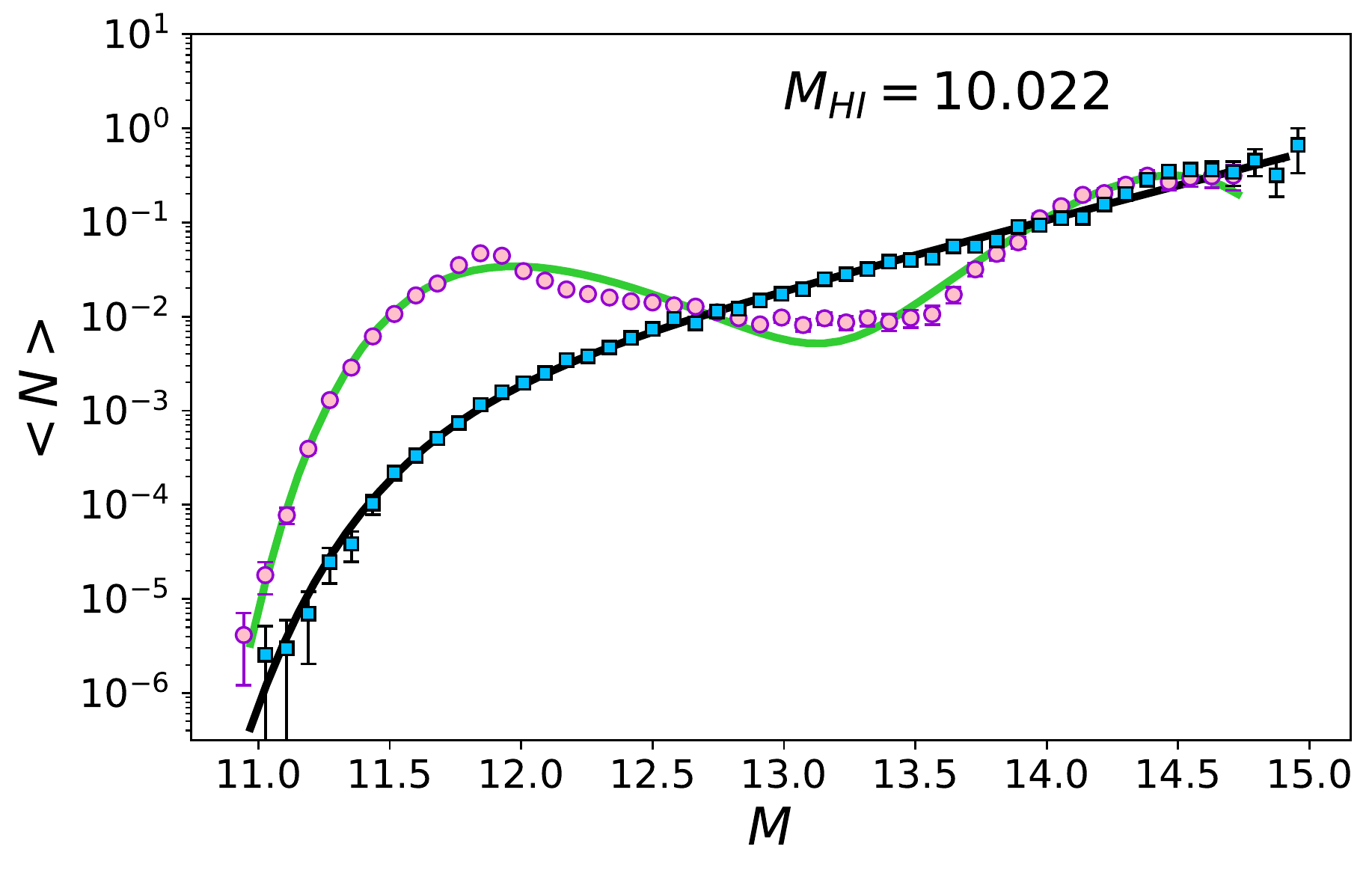}
 \caption{Fits of our model HOD, Eq.~\ref{HODcen6} and \ref{HODsat}, to the measured HOD. Three example $M_{\rm HI}$ bins are shown. The black and green curves are the best-fit satellite and central models respectively, while the blue-squares and pink-circles show the measurements from these two classes of galaxy.}
 \label{HItoHODapdix}
\end{figure}

Before fitting the H\thinspace{\protect\scriptsize I}-conditional HOD models using the galaxy two-point correlation function, we need to explore how well the functional forms of Eq.~\ref{HODcen6} and \ref{HODsat} compare to the measured HOD in different mass bins. Therefore, we directly fit Eq.~\ref{HODcen6} and \ref{HODsat} to the measured HOD. To measure the HOD, we divide the halo mass into 50 bins in the interval $M\in[10.9,15]$. Then in each halo mass bin, we divide the H\thinspace{\protect\scriptsize I}  mass in the interval $M_{\rm HI}\in[9.11,10.3]$ with bin width $dM_{\rm HI}=0.079$. The fit results for three example mass bins are shown in Fig.~\ref{HItoHODapdix}. Our model curves can fit the central and satellite HODs remarkably well over a range of H\thinspace{\protect\scriptsize I} and halo mass bins, which indicates our HOD model is flexible enough to fit the data.

\subsection{Reducing the parameters of the H\thinspace{\protect\scriptsize I}-conditional HOD model of the central galaxies}
\label{sec:resultsimplify}

When fitting the HOD model to the various mass bins, we are introducing a large number of free parameters (9 per H\thinspace{\protect\scriptsize I} mass bin, so a total of 135 over the entire simulation). This could present a problem when fitting the clustering, due to difficulties optimising in such a high-dimension parameter space, which introduces a high chance of overfitting. So, initially, we investigate whether there are any trends in the best-fit HOD parameters between H\thinspace{\protect\scriptsize I} mass bins that can enable us to simplify the model.

\begin{figure} 
\centering
\includegraphics[width=\columnwidth]{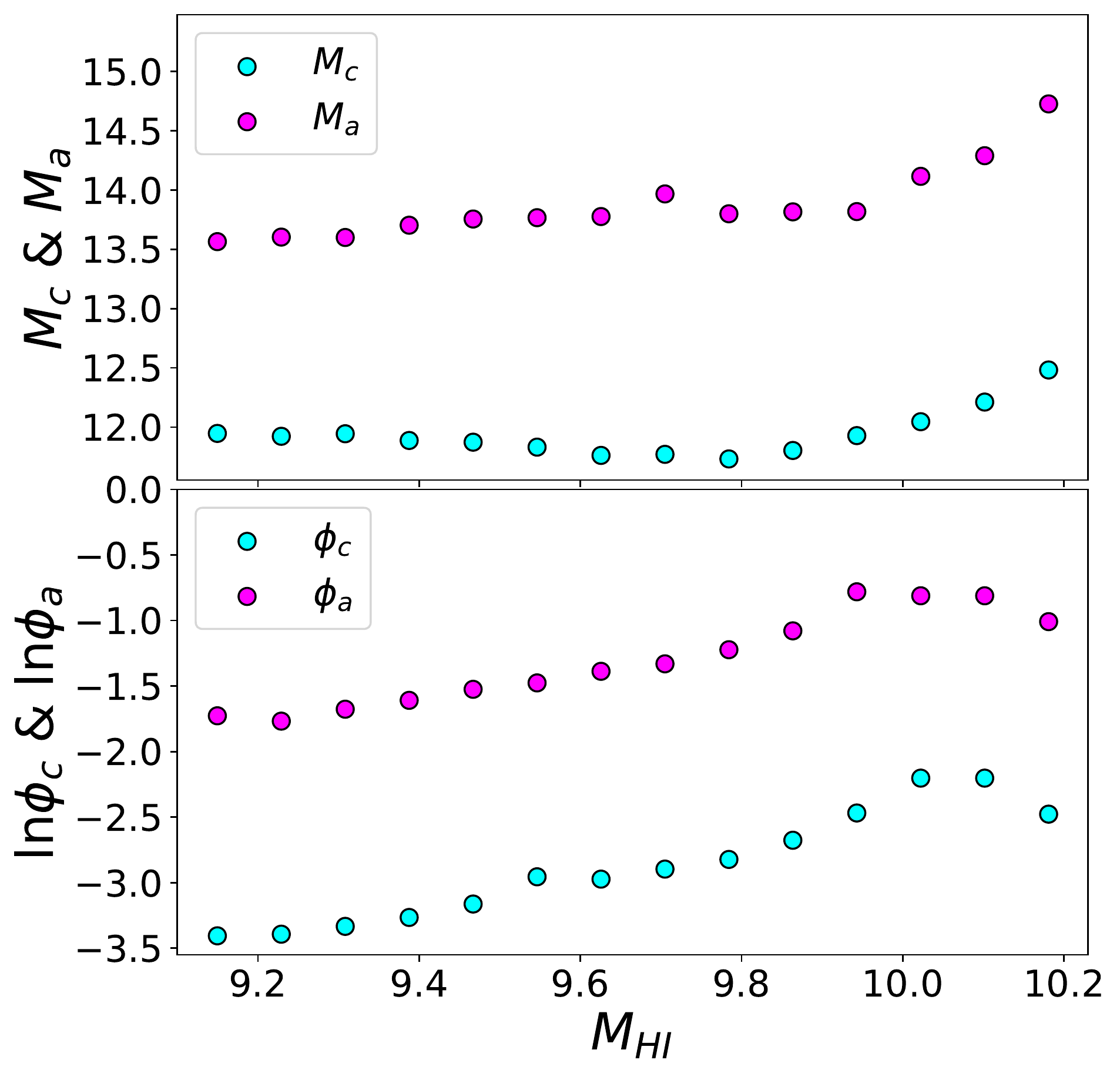}
 \caption{ The best-fit values of the HOD parameters $M_c$, $M_a$, $\phi_c$ and $\phi_a$ as functions of H\thinspace{\protect\scriptsize I} mass.}
 \label{Pam9s}
\end{figure}

Firstly, we find particularly interesting correlations between two pairs of parameters shown in Fig.~\ref{Pam9s}. For one, we find that $M_a$ shows an almost constant shift compared to $M_c$. We also then see that $\ln\phi_a$ shows an approximately constant shift compared to $\ln\phi_c$ too. These pairs of parameters are the locations and relative amplitudes of the two `knees' of the double Schechter function used to parameterise the central galaxy HOD.

\begin{figure} 
\centering
\includegraphics[width=\columnwidth]{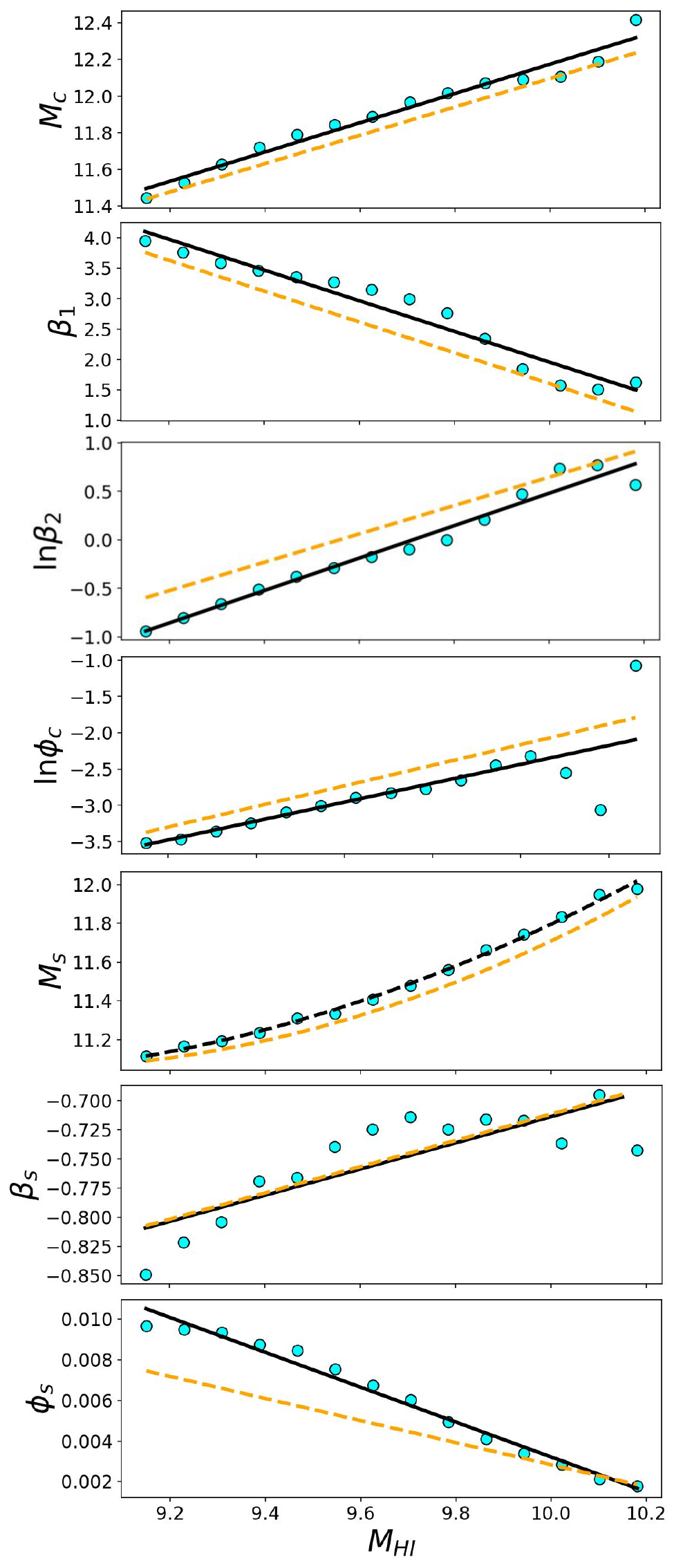}
 \caption{ The parameters of Eq.~\ref{HODcen6} and \ref{HODsat} as functions of H\thinspace{\protect\scriptsize I} mass.  The blue filled circles are the parameters fitted by comparing the model HOD to the measured HOD in each H\thinspace{\protect\scriptsize I} mass bin. The black lines and curve are the best fits to the blue filled circles, resulting in our 15 `hyper-parameters'. The orange dashed lines and curve are the best-fit results obtained by comparing the model $\xi_{gg}$ to the measured $\xi_{gg}$.  Differences between this figure and Fig.~\ref{Pam9s} for overlapping parameters arise due to the fact that we have forced the remaining two of the parameters of the full HOD model (not plotted here) to follow Eq.~\ref{MPHIsimp}, which was not the case for Fig.~\ref{Pam9s}.}
 \label{HODHIm}
\end{figure}

Given the relation between these pairs, we can reduce the number of parameters of
Eq.~\ref{HODcen6} by setting
\be \label{MPHIsimp}
M_a = M_c+2~,~~\phi_a= 5\phi_c.
\ee 
We are then left with only 7 free parameters per H\thinspace{\protect\scriptsize I} mass bin (or 105 for the full simulation). For these remaining parameters, we plot the fit results against the mean H\thinspace{\protect\scriptsize I} mass in each bin in Fig.~\ref{HODHIm}.

Looking at these, we see that they nearly all present as (close to) linear or quadratic functions of the H\thinspace{\protect\scriptsize I} mass. We hence implement our H\thinspace{\protect\scriptsize I} mass conditional HOD by enforcing relationships between the parameters ${\bf{P}}=\{M_c,  \beta_1, \ln\beta_2, \ln\phi_c,\beta_s,\phi_s\}$ and $M_{\rm HI}$ using linear functions,
\be \label{PvsHi}
{\bf{P}}_i=k_iM_{\rm HI}+b_i~.
\ee
While $M_s$ is a quadratic function of H\thinspace{\protect\scriptsize I} mass,
\be  \label{PvsHi22}
M_s=c_0+c_1M_{\rm HI} +c_2M_{\rm HI}^2~,
\ee
Fitting our HOD parameters as a function of H\thinspace{\protect\scriptsize I} mass, we recover the black lines in Fig.~\ref{HODHIm}. 
The best-fit values of $(k_i,~b_i)$ are given in the first two columns of Table \ref{bkflb} while $c_0=42.748 ,~ c_1=-7.352 ,~ c_2=0.426$.

\begin{table}   \small
\centering 
\caption{ The best-fit values of $k$ and $b$ from Eq.~\ref{PvsHi} for each relevant HOD parameter both from fitting the measured HODs in each H\thinspace{\protect\scriptsize I} mass bin directly (Section~\ref{sec:resultsimplify}) and from fitting the galaxy correlation functions (Section~\ref{secHIHODss}).}
\begin{tabular}{|c|c|c|c|c|}
\hline
\hline

\multirow{2}*{${\bf{P}}$} &\multicolumn{2}{|c|}{Fit from the measured HOD} &\multicolumn{2}{|c|}{Fit from $\xi_{gg}$} \\
\cline{2-5}
       & $k_i$ & $b_i$ & $k_i$ & $b_i$ \\
\hline
  $M_c$ & $0.799$ & $4.183$  & $0.774$ & $4.353$  \\ 
\hline
  $\beta_1$  & $-2.528$ & $27.234$ & $-2.538$ & $26.981$ \\
\hline
  $\ln\beta_2$   & $1.678$ & $-16.295$ & $1.459$ & $-13.948$\\
\hline
  $\ln\phi_c $    & $1.302$ & $-15.453$ & $1.419$ & $-16.353$\\
\hline
  $\beta_s$& $0.112$ & $-1.838$ & $0.112$ & $-1.839$\\
\hline
  $\phi_s $  & $-0.00858$ & $0.0890$ & $-0.00544$ & $0.0572$ \\
\hline
\end{tabular}
 \label{bkflb}
\end{table}

\begin{figure} 
\centering
\includegraphics[width=\columnwidth]{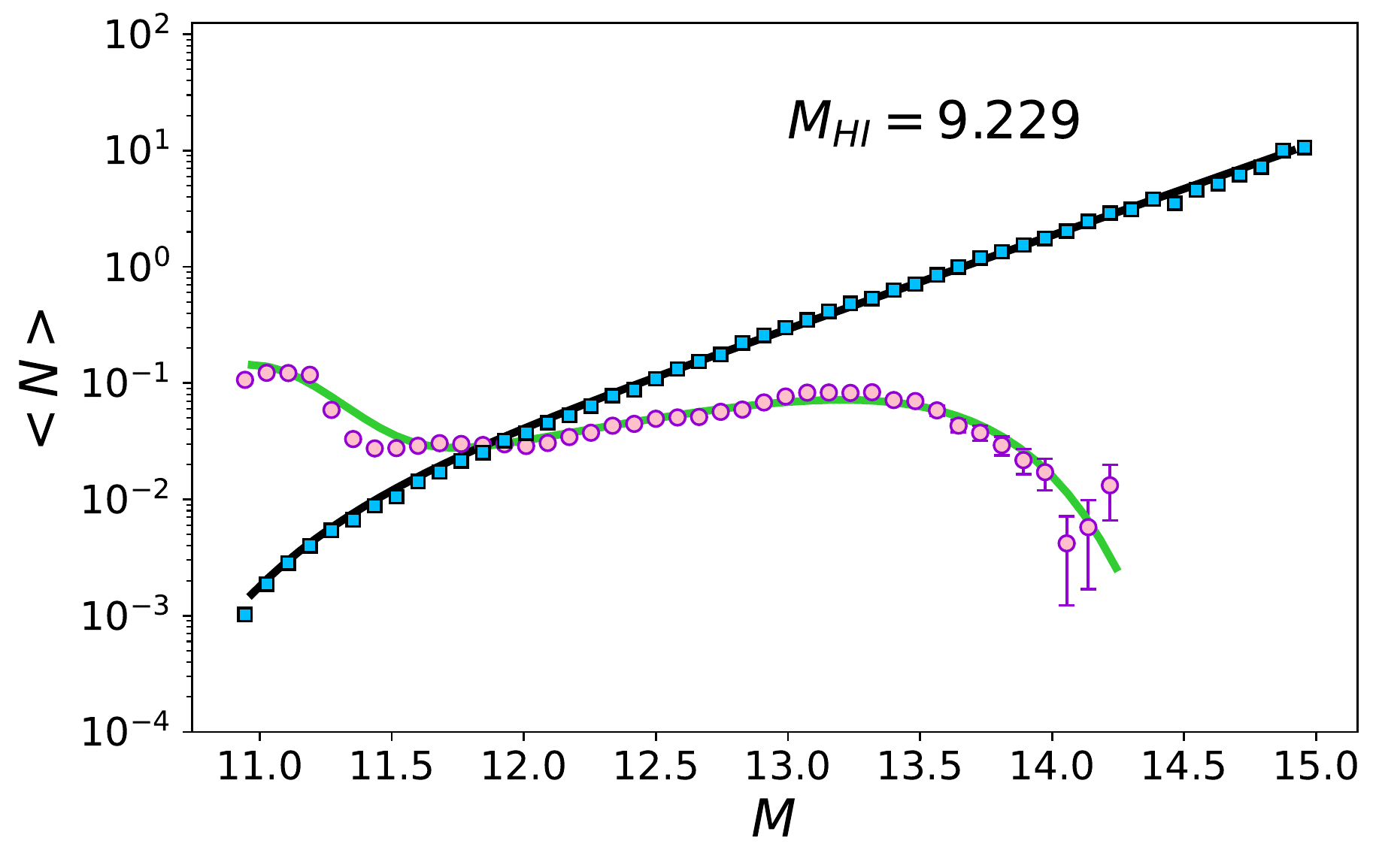}
 \includegraphics[width=\columnwidth]{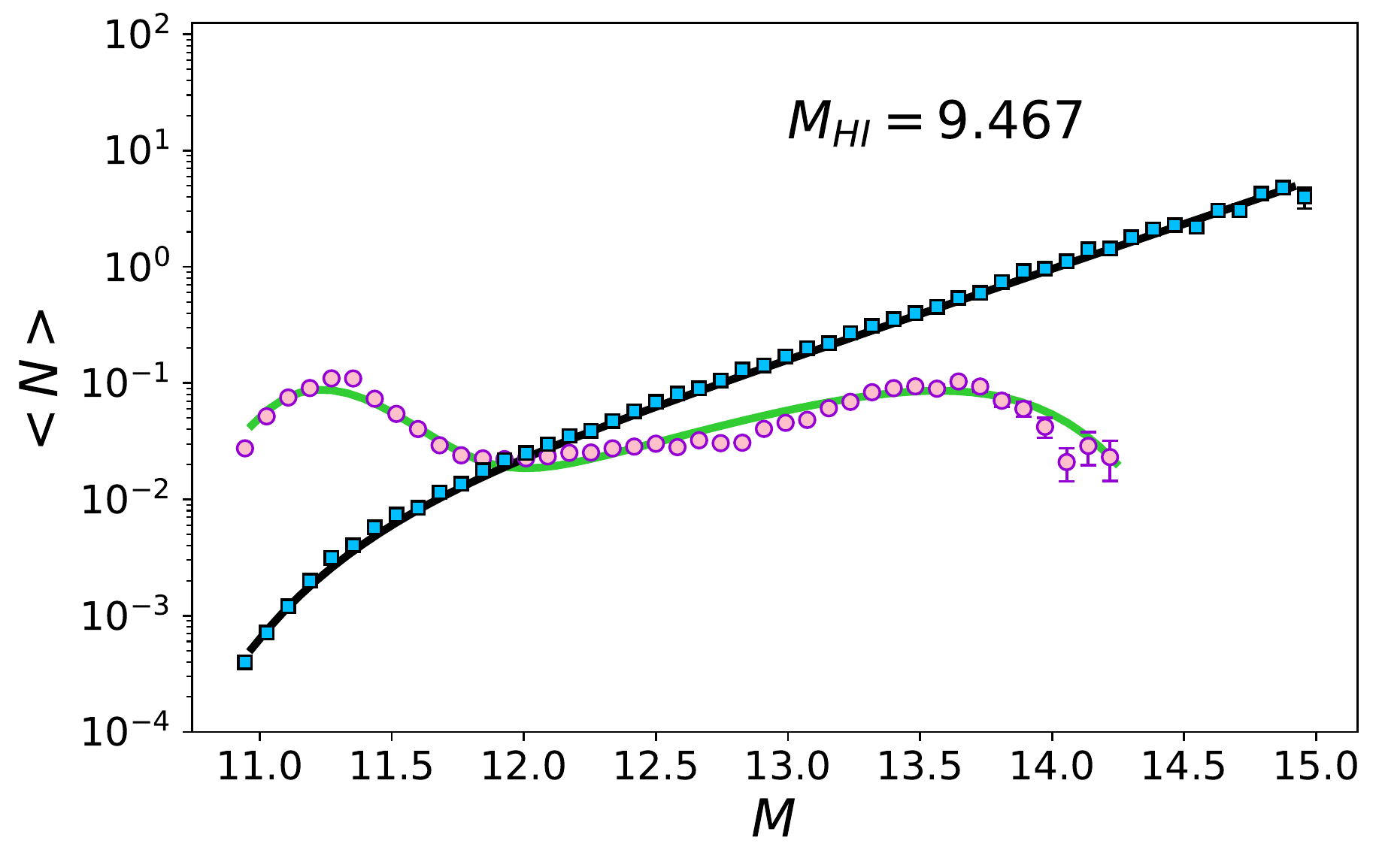}
 \includegraphics[width=\columnwidth]{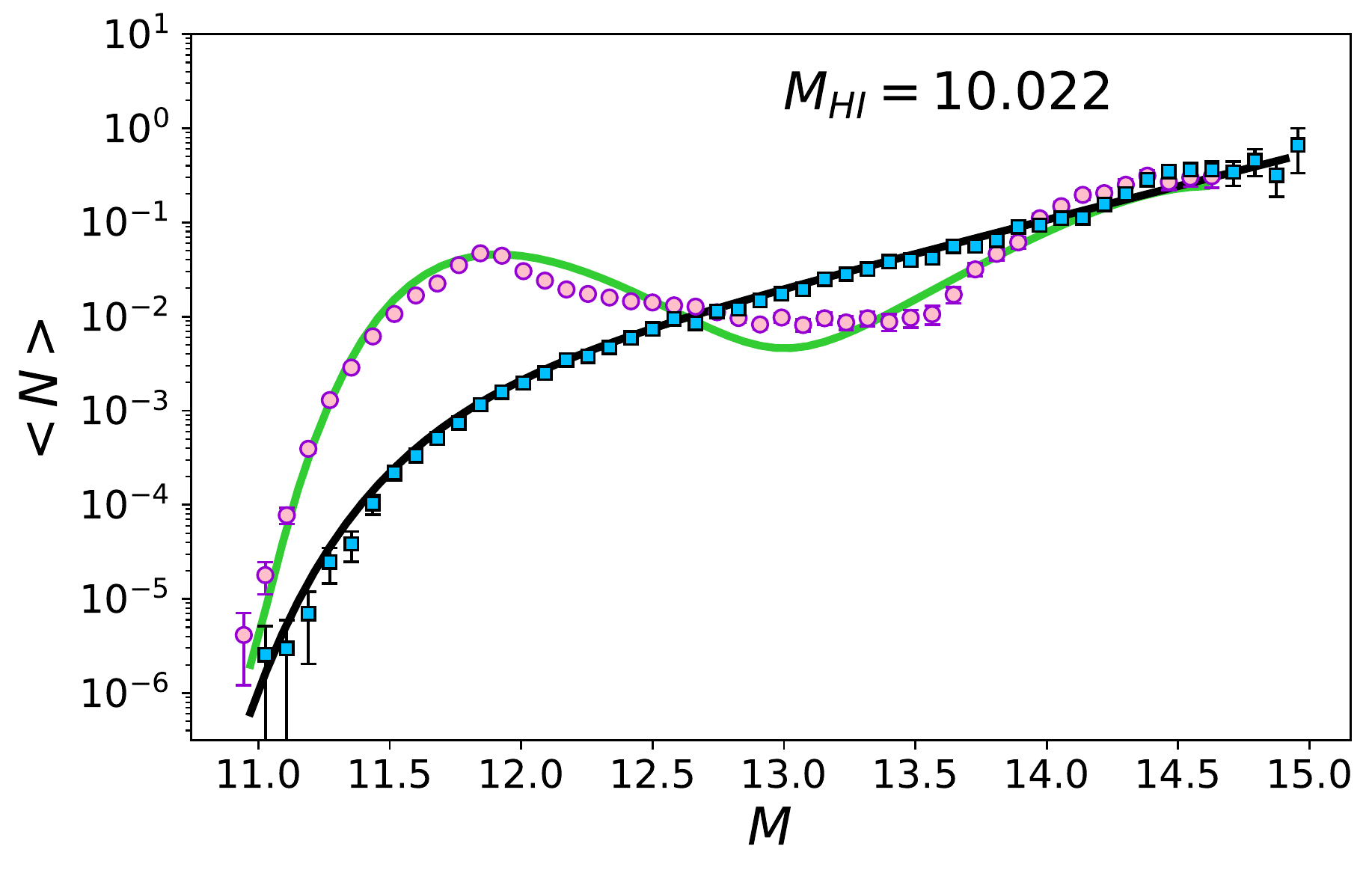}
 \caption{Comparison of the model HOD (green and black curves for central and satellite galaxies respectively) predicted by our 15 best-fit `hyper-parameters' and the measured HOD (pink-circles and blue-squares for centrals and satellites respectively) for three example $M_{\rm HI}$ bins. This plot looks functionally very similar to Fig.~\ref{HItoHODapdix}, which emphasises the fact that we are able to recover well the measured HODs for various H\thinspace{\protect\scriptsize I} mass bins using our `hyper-parameters', rather than fitting each bin individually.}
 \label{HItoHOD}
\end{figure}

With the above procedure, we have reduced the full set of 135 HOD parameters in our model down to only 15 `hyper-parameters'. Our parameterisation directly maps halo mass to the number of galaxies \textit{of a given H\thinspace{\protect\scriptsize I} mass}, and so can be used to populate halos not just with a number of galaxies (as in a standard HOD), but assign H\thinspace{\protect\scriptsize I} masses to those galaxies too. For example, plugging $M_{\rm HI}=9.229$ into Eq.~\ref{PvsHi} and \ref{PvsHi22} one can obtain the HOD parameters $M_c$, $\beta_1$, $\beta_2$, $\phi_c$ and $M_s$, $\beta_s$, $\phi_s$. These can then be plugged into Eqs.~\ref{MPHIsimp}, ~\ref{HODcen6} and~\ref{HODsat} to obtain the HOD of central and satellite galaxies, with the results shown in the top panel of Fig.~\ref{HItoHOD}. Another two example H\thinspace{\protect\scriptsize I} mass bins are also shown in Fig.~\ref{HItoHOD}. Overall, after exploring all H\thinspace{\protect\scriptsize I} mass bins, we find that the HOD calculated from the H\thinspace{\protect\scriptsize I} mass using this procedure still matches the measured HOD exceptionally well despite the large reduction in the number of free parameters we need.

We caution that the exact values of the $(k_i,~b_i)$ and $c_0 ,~ c_1,~ c_2$, and the constant relations seen in Eq.~\ref{MPHIsimp} are specific to the {\sc Dark Sage} simulation. However, we believe that enough flexibility remains for the same model to fit real data, given that the simulation reproduces the $z=0$ H\thinspace{\protect\scriptsize I} mass function, stellar mass function and H\thinspace{\protect\scriptsize I} mass-to-stellar mass ratio of the real Universe \citep{Stevens2018}. It would hence be interesting to explore if this is the case, and whether changes to the exact functional forms described above are needed. However, we leave this for the next work in the series, and for the remainder of this paper demonstrate that the conditional HOD described above can also be constrained purely from measurements of the galaxy clustering, without requiring prior knowledge of the halo masses themselves.

\subsection{Fitting the H\thinspace{\protect\scriptsize I}-conditional HOD using the galaxy two-point correlation function}\label{secHIHODss}

We test our model further by fitting the measured two-point correlation functions for the galaxies in each H\thinspace{\protect\scriptsize I} mass bin from Section~\ref{sec:ResultHOD}, concatenating them together to obtain a single data vector. For a given set of HOD parameters $k_i$, $b_i$, $c_0,~ c_1,~ c_2$, we calculate the HOD models for each H\thinspace{\protect\scriptsize I} mass bin using Eqs.~\ref{PvsHi}, \ref{PvsHi22}, \ref{MPHIsimp}, \ref{HODcen6} and \ref{HODsat}. We then calculate the model correlation function in each H\thinspace{\protect\scriptsize I} mass bin using the methodology of Section~\ref{sec:Xigg} and measured radial profiles for the simulation shown in Section~\ref{sec:radialprofile}, before again concatenating them together to obtain a single model vector. Finally, we minimize the difference between the data and model to find the best-fitting $k_i$, $b_i$, $c_0,~ c_1,~ c_2$.

In this case, we are using the least squares difference rather than $\chi^2$ minimization to fit the model correlation functions to the measurements since it is not possible to estimate the full covariance matrix of the correlation function using a single {\sc Dark Sage} simulation and jackknife. There are over 500 correlated measurement bins, far larger than the number of jackknife samples we could generate, leading to a noisy and ill-conditioned estimate of the covariance matrix. In future work, we will develop a more robust method for the fitting that iterates between populating an ensemble of simulations with an HOD, using this ensemble to estimate the covariance matrix, and then fitting the data to generate the next HOD with which to populate the ensemble (i.e., as done in \citealt{Howlett2015a, Howlett2022, Qin2019b, Qin2021b}). Nonetheless, the methodology used here is sufficient for the purposes of this proof-of-concept  study.

 The above fitting method can be applied to real galaxy surveys too. For example, in the WALLABY data, the \HI~masses of hundreds of thousands of galaxies will be measured in the mass range we have studied here \citep{Koribalski2020}. Therefore, the two-point correlation functions of the galaxies in each \HI~mass bin can be measured and then fitted to the model correlation functions to estimate the HOD parameters in the same fashion as the simulation data. This is in fact crucial to understanding the relationship between \HI~galaxies and their host halos, as if we were to measure and fit only a single correlation across all WALLABY galaxies, we would be only able to constrain the total \HI-to-halo mass ratio (i.e., as in \citealt{Obuljen2019} --- see Appendix \ref{sec:hihmr}), rather than how that total \HI~mass is distributed across individual central and satellite galaxies.

\begin{figure*} 
\centering
\includegraphics[width=59mm]{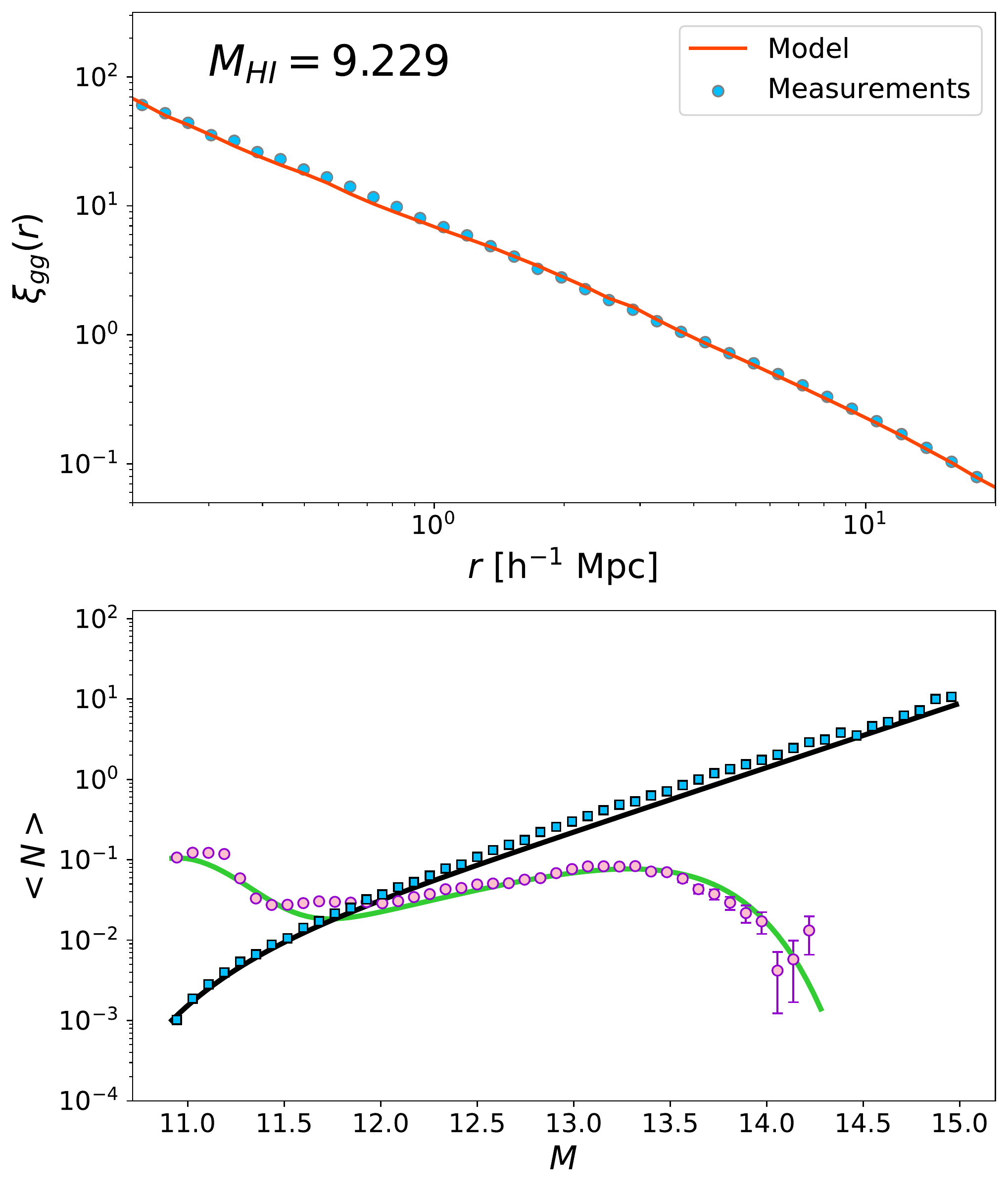}
 \includegraphics[width=59mm]{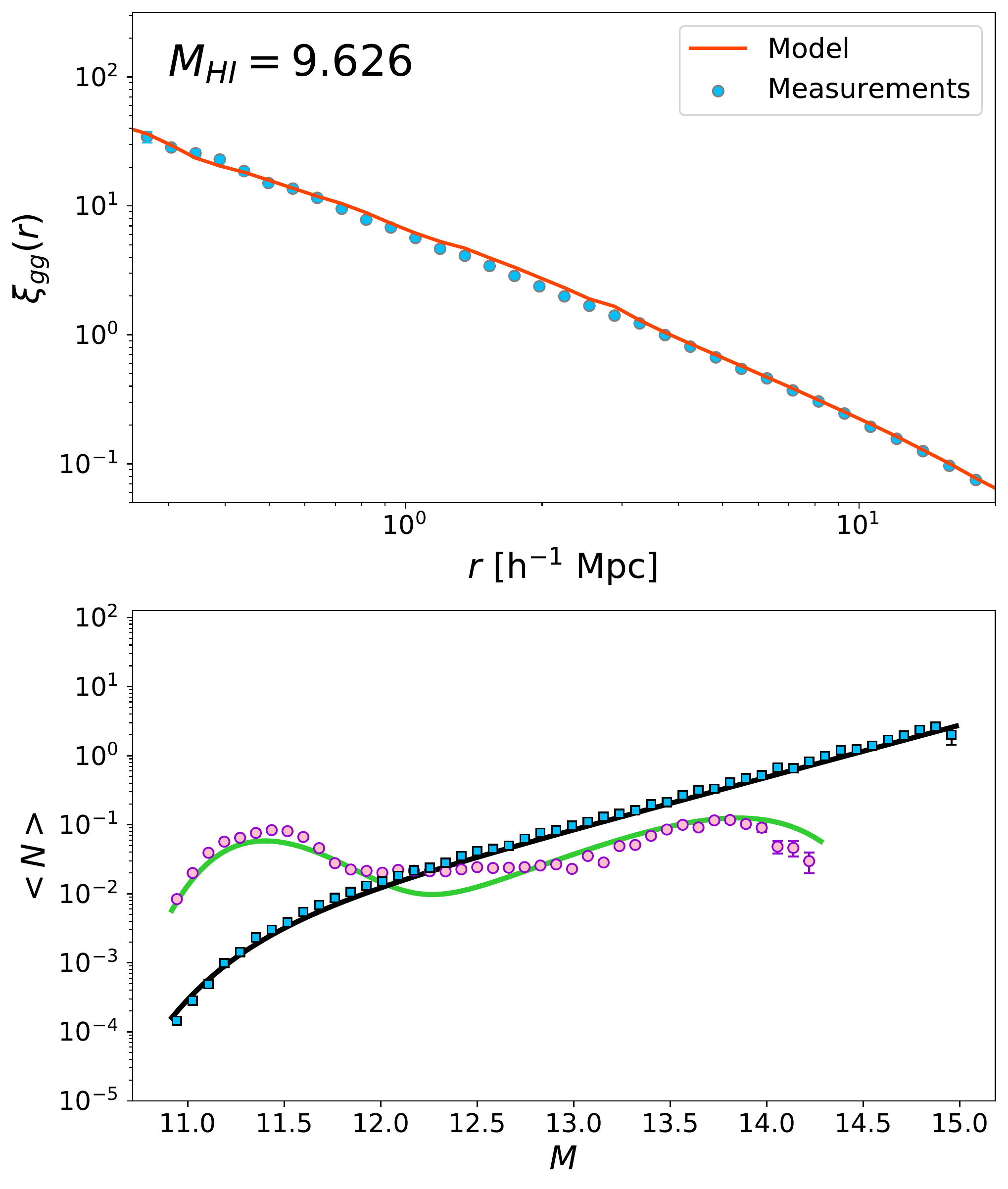}
 \includegraphics[width=59mm]{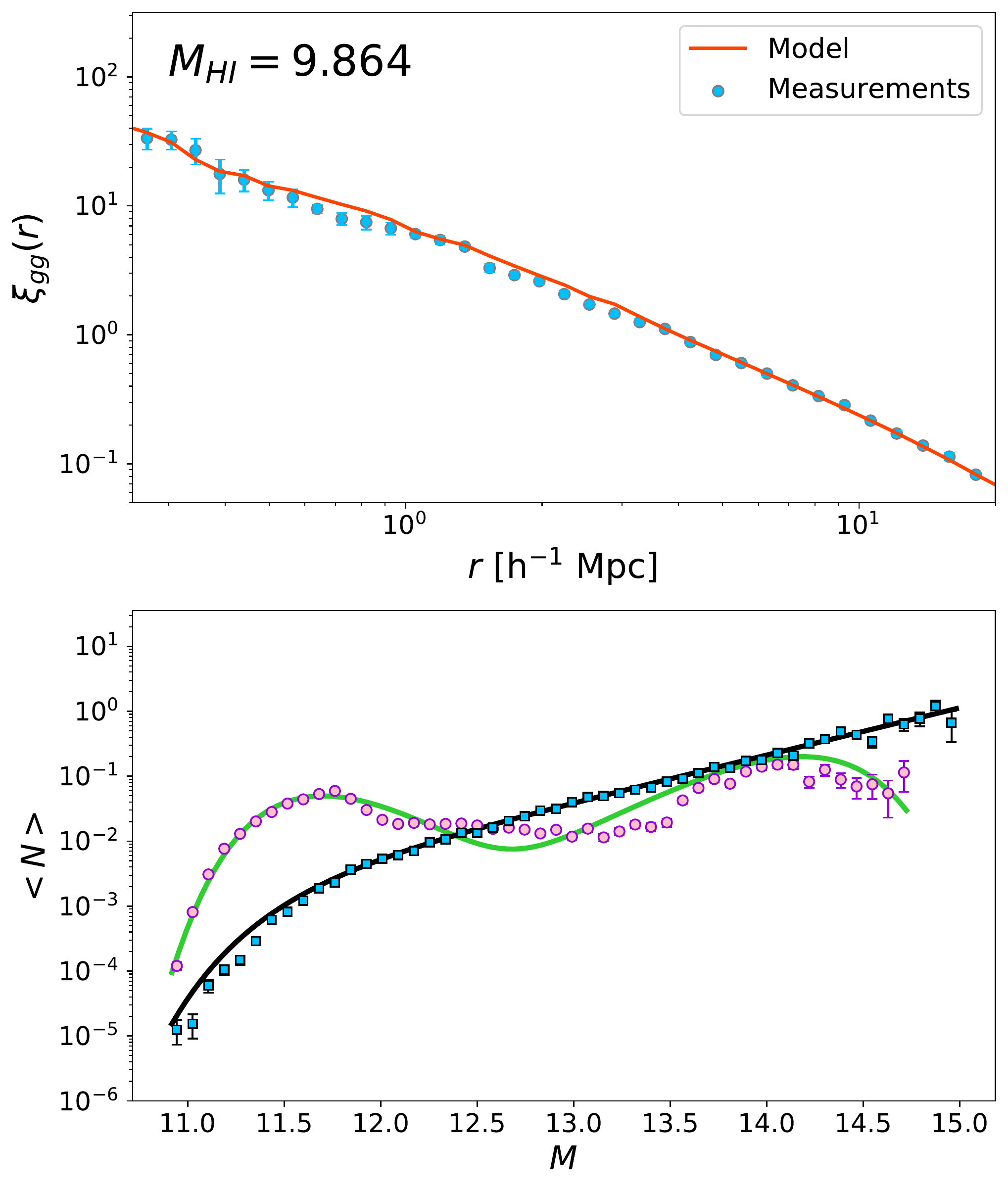}
 \caption{A demonstration of fitting the HOD of H\thinspace{\protect\scriptsize I} galaxies using the galaxy two-point correlation function. Three example $M_{\rm HI}$ bins are shown. The panels in the top row show the measured and modelled correlation functions (blue circles and red curves respectively). The bottom panels show the measured HOD for centrals and satellites (pink-circles and blue-squares respectively) against that predicted (green/black curves for centrals/satellites) by our best-fitting set of `hyper-parameters' obtained from fitting the correlation functions.}
 \label{XiggHODss}
\end{figure*}

In Fig.~\ref{XiggHODss} we show the results from simultaneously fitting all 15 correlation functions with our conditional HOD model containing 15 free parameters. Although we are unable to report a chi-squared for the fit, we find models for the correlation functions that agree well visually across all the scales and H\thinspace{\protect\scriptsize I} mass bins we consider. More remarkable is the agreement seen in the lower panels of this figure between the measured HOD of each H\thinspace{\protect\scriptsize I} mass bin and that \textit{predicted} from fitting the correlation functions. We see model HODs that align well with the measurements, despite us having the equivalent of only a single parameter per bin. We reiterate that we have not allowed the HOD parameters in each bin to vary independently, but rather are forcing the parameters describing the shapes of the various Schechter functions to themselves depend only linearly or quadratically on the value of $M_{\rm HI}$ for a given bin. We hence are able to fit the many correlation functions without any knowledge of the halo masses themselves, only the mean H\thinspace{\protect\scriptsize I} mass.

The recovered linear and quadratic functions of H\thinspace{\protect\scriptsize I} mass are shown as the dashed-orange lines and curve in Fig.~\ref{HODHIm}. The best estimated values of $k_i$, $b_i$ are listed in the last two columns of Table \ref{bkflb}, while $c_0=51.321 ,~ c_1=-9.084 ,~ c_2=0.512$. Here, we do see some differences between the parameters obtained from directly fitting the measured HOD, and those \textit{inferred} from fitting the correlation functions. These arise primarily due to 1) degeneracies across the full parameter space, which mean that multiple values of our fitting parameters can return similar correlation functions; 2) noise in the measured $\xi_{gg}$, at large separations (due to the finite volume of the simulation), or at small separation/bins with larger H\thinspace{\protect\scriptsize I} mass (due to shot noise); and 3) noise in the \textit{model} $\xi_{gg}$, especially in larger H\thinspace{\protect\scriptsize I} mass bins with fewer galaxies, due to the use of measured radial distributions for the simulated halos. Nevertheless, our method is still able to reproduce the clustering of the H\thinspace{\protect\scriptsize I} galaxies well.

\begin{figure} 
\centering
\includegraphics[width=\columnwidth]{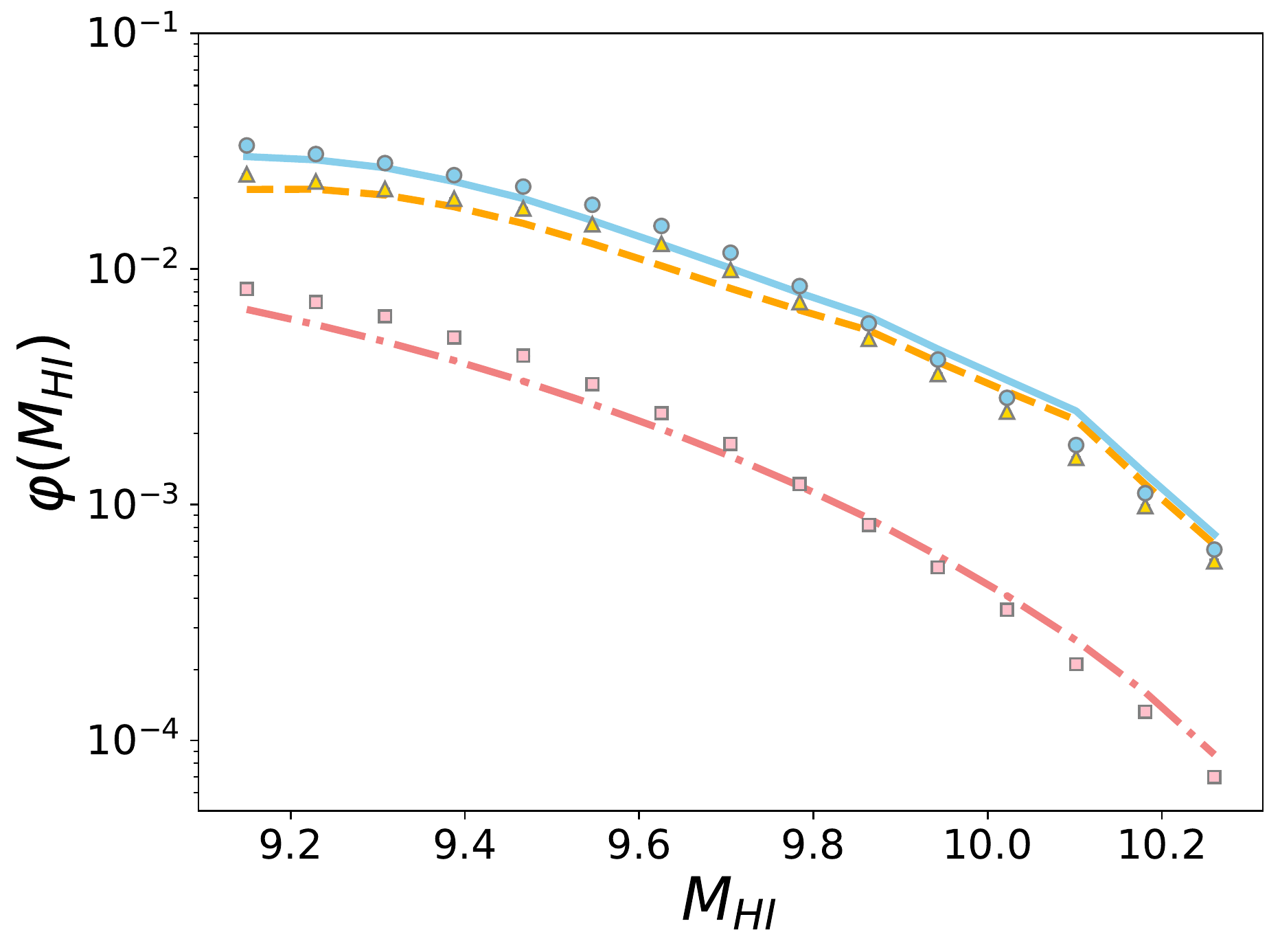}
 \caption{ The measured H\thinspace{\protect\scriptsize I} mass function of {\sc Dark Sage} (points) compared to the mass function predicted (lines) by integrating over our best-fit conditional HOD (with parameters given in the last two columns of Table~\ref{bkflb}). Blue points and curves are the total mass function, which we further split into centrals (yellow) and satellites (pink). The error-bars are tiny.}
 \label{HIMF_HODs}
\end{figure}

One very interesting by-product of fitting the H\thinspace{\protect\scriptsize I}-conditional HOD using the galaxy clustering is that we can take an HOD model and integrate over halo mass to recover a prediction for the H\thinspace{\protect\scriptsize I} mass function of galaxies. As a final check of our method we do this for our best-fit above, and compare to the H\thinspace{\protect\scriptsize I} mass function measured from the {\sc Dark Sage} simulation. This is shown in Fig.~\ref{HIMF_HODs}. Again, despite not including the H\thinspace{\protect\scriptsize I} mass function in our fit, we find good agreement between the measurements and predictions, both for all galaxies and when split into separate central and satellite populations. We note that this agreement could be made even better by including the observed H\thinspace{\protect\scriptsize I} mass function alongside the correlation functions in our joint data vector, which we anticipate would break some of the degeneracies between our `hyper-parameters' and result in greater agreement between the measurements and predictions in Fig.~\ref{HODHIm}. However, we find it more revealing to leave this as a cross-validation for now, and defer including it in the fit to future work.
 
\section{Conclusion} \label{sec:conc}

In this paper, we have presented a new HOD model for H\thinspace{\protect\scriptsize I} galaxies, using single and double Schechter functions that predict the number of central and satellite galaxies (Eqs.~\ref{HODcen6} and \ref{HODsat}) in a halo, and functions of the neutral hydrogen  mass ($M_{\rm HI}$) that predict the parameters of these Schechter functions (encapsulated by \ref{MPHIsimp}, \ref{PvsHi} and \ref{PvsHi22}). In much the same way as the conditional luminosity or stellar mass functions \citep{Cooray2006, Guo2018}, this enables us to predict the number of galaxies of a given H\thinspace{\protect\scriptsize I} mass in halos of a given halo mass, and so produce mock catalogues that contain H\thinspace{\protect\scriptsize I} masses in addition to positions and velocities. This will be crucial for constraining cosmology from upcoming radio surveys such as WALLABY \citep{Koribalski2020} or with the SKA \citep{2020PASA...37....7S}.

The functional form used for the HOD model contains a total of 15 free parameters, and is determined from comparison with the {\sc Dark Sage} simulation. We have demonstrated that these parameters can be fit well to the galaxy correlation function measured in H\thinspace{\protect\scriptsize I} mass bins and that applying this to the simulation results in good agreements between the measured and predicted conditional HOD, and H\thinspace{\protect\scriptsize I} mass function. This is encouraging, as it means one could consider making such measurements with real data, fitting these to obtain the best-fit HOD parameters, and then using these to populate simulations in a way that reproduces the clustering and H\thinspace{\protect\scriptsize I} mass function of the data. 

One caveat to our work is that we found the model to be very sensitive to the assumed radial distribution of satellite galaxies in their host halos, as any HOD model would be. We measured the radial profile of the H\thinspace{\protect\scriptsize I} galaxies in the {\sc Dark Sage} simulation and found that they do not follow the NFW profile. Hence, accurately calculating the model correlation function requires us to use the measured radial profiles rather than analytic.

There are many extensions to this first work that can be pursued, including further investigation of the impact of the halo radial profiles on the conditional HOD, the inclusion of H\thinspace{\protect\scriptsize I} mass function as data to constrain the HOD parameters, and improvements to the overall robustness of the fitting method we use here (e.g., properly accounting for the covariance between correlation function measurements).  We also restricted our analysis to real-space positions of galaxies measured from the simulation. In future work, we aim to improve on these and develop a complete mock sampling algorithm for H\thinspace{\protect\scriptsize I} surveys that can be used to generate mocks, for example for the redshift and Tully--Fisher catalogues of WALLABY \citep{Koribalski2020}.  For example, we anticipate that a simple extension would be to use a lightcone simulation with galaxy redshifts to measure and model the redshift-space clustering by also drawing radial velocities (from the NFW profile, or other suitable distribution) alongside positions when populating the halos with galaxies (as in \citealt{Howlett2015,Howlett2022,Qin2019b,Qin2021b}).

\acknowledgments

FQ and DP are supported by the project \begin{CJK}{UTF8}{mj}우주거대구조를 이용한 암흑우주 연구\end{CJK} (``Understanding Dark Universe Using Large Scale Structure of the Universe''), funded by the Ministry of Science. CH is supported by the Australian Government through the Australian Research Council’s Laureate Fellowship funding scheme (project FL180100168).
ARHS is funded through the Jim Buckee Fellowship at ICRAR-UWA.

%



\software{
          \textsc{CAMB} \citep{Lewis:1999bs},
          \textsc{ChainConsumer} \citep{ChainConsumer}, 
          \textsc{Corrfunc}  \citep{Corrfunc2020},
          \textsc{Dark Sage} \citep{Stevens2017code},
            \textsc{emcee} \citep{Foreman-Mackey2013}, 
          \textsc{Matplotlib} \citep{Hunter2007},
          \textsc{Nbodykit} 
          \citep{Nbodykit2018},  
                    \textsc{Scipy} \citep{Virtanen2020}.
}

\appendix

\section{The halo biasing parameter}\label{sec:bh}

The halo biasing parameter $b_h(M)$ is defined as \citep{Tinker2005}
 \be \label{bhvs}
 b_h^2(M)\equiv\frac{\xi_h(r,~M)}{\xi_m(r)}~,
 \ee  
 where $\xi_h(r,~M)$ is the two-point correlation function of the halos of mass $M$, while $\xi_m(r)$ is the {\it nonlinear} matter correlation function. We use the \verb~CAMB~ package to generate the nonlinear matter power spectrum, then convert it into the nonlinear matter correlation function\footnote{We use the function cosmology.pk\_to\_xi of the \verb~PYTHON~ package \verb~Nbodykit~. \url{https://nbodykit.readthedocs.io/en/latest/api/_autosummary/nbodykit.cosmology.correlation.html}.}.
 
\begin{figure} 
\centering
 \includegraphics[width=\columnwidth]{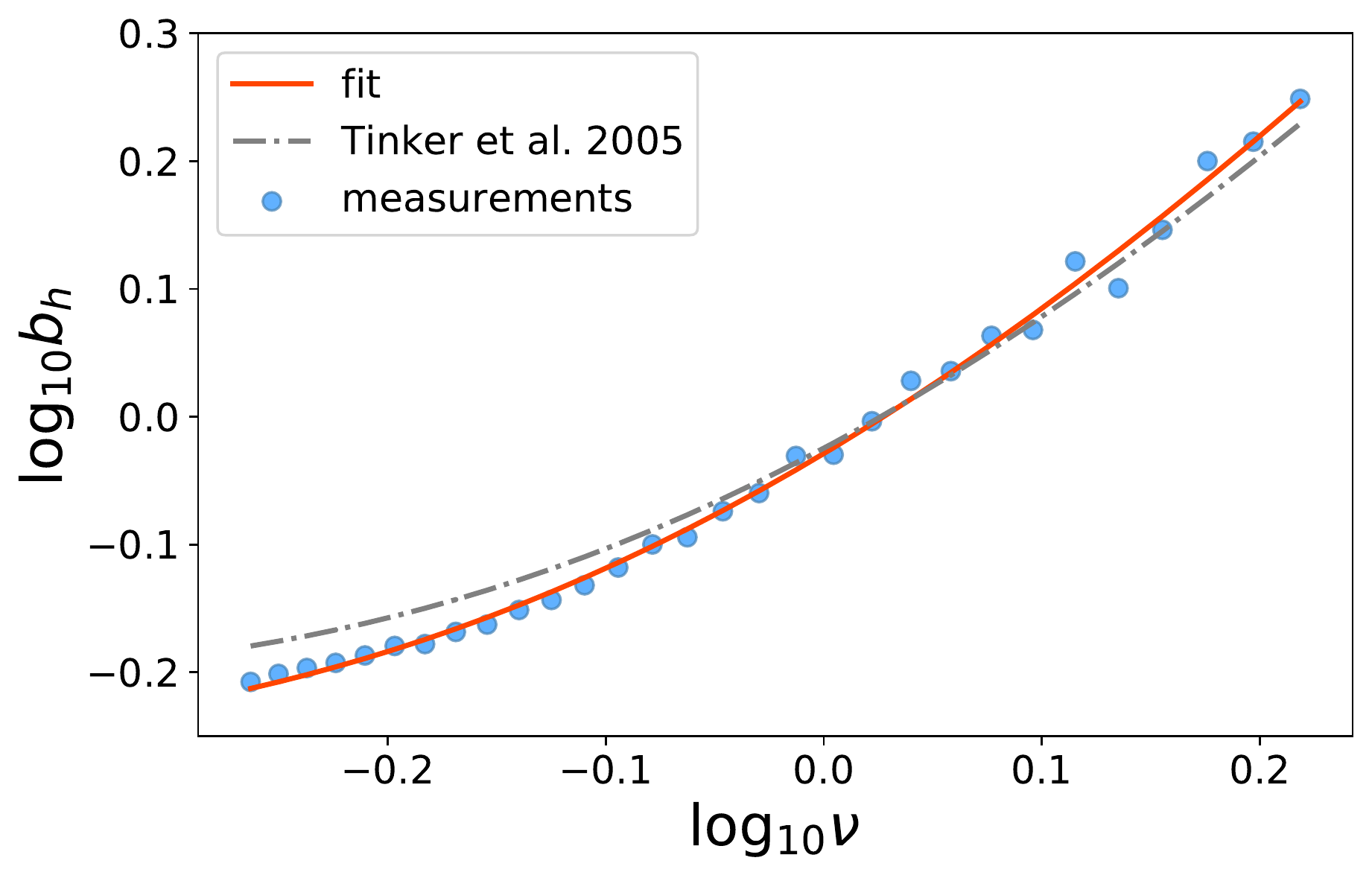}
 \caption{The halo biasing parameter as a function of $\nu= \delta_c / \sigma(m)$. The blue dots are the measured halo biasing parameters of the {\sc Dark Sage} simulation, the red curve is the model fit to the measurements, and the grey curve is the fit from \cite{Tinker2005}.}
 \label{bhv}
\end{figure}

Following the arguments in appendix A of \citet{Tinker2005}, to measure the halo biasing parameter of {\sc Dark Sage} simulation, we firstly divide the parent halos into 30 mass bins, then measure $\xi_h(r,~M)$ of the halos in each mass bin. Finally, we calculate $b_h$ in each halo mass bin by averaging $\sqrt{\xi_h/\xi_m}$ for pair separations $4\leq r\,h/{\rm Mpc} \leq15$. In this interval, the ratio between halo and matter correlation function is approximately constant and the measured correlation functions are smooth. In Fig.~\ref{bhv}, the blue dots represent the measured halo biasing parameters as a function of 
  $\nu=\delta_c / \sigma(m)$, where the critical density contrast for collapse is $\delta_c=1.686$.

 The model halo biasing parameter can be written as \citep{Sheth2001b,Yang2003,Tinker2005,Zheng2007}
\be \label{bhvm}
\begin{split}
 b_h(\nu)=&1+\frac{1}{\sqrt{a}\delta_c}\times  \bigg  [ \sqrt{a} (a\nu^2) +\sqrt{a}b(a\nu^2)^{1-q}   \\
   - & \frac{(a\nu^2)^q}{(a\nu^2)^q+b(1-q)(1-q/2)} \bigg ] ~.
\end{split}
\ee 
Fitting the model to the measurements, we find $a=0.79168$, $b=0.21136$ and $q=0.66108$ for {\sc Dark Sage}, which differs slightly from the values found in \citep{Tinker2005}. The measurements, our best-fit model, and the \cite{Tinker2005} model are shown in Fig.~\ref{bhv}.

\section{NFW Halo Density Profile}\label{sec:nfw}

A commonly used halo density profile for the matter is the Navarro--Frenk--White (NFW) profile \citep{NFW1996,Navarro1997}, and the  
radial distribution of mass density is given by \citep{NFW1996,Navarro1997,Yang2003,Zheng2007}
\be\label{NFWden}
\rho_m(R)=\frac{\rho_s}{(R/R_s)(1+R/R_s)^2}~,
\ee 
where
\be \label{NFWdencdcd}
\rho_s=\frac{\rho_0\Delta}{3}\frac{c^3}{\ln(1+c)-c/(1+c)}~,
\ee  
and the halo concentration is defined as
\be 
c\equiv \frac{R_{\mathrm{vir}}}{R_s}~.
\ee 
$R_s$ is the  break  radius  between  the  outer and inner density  profile of the halo.
We use the fitting formulae of \citet[][section 5 therein]{Prada2012} to calculate $c$ for the halos as a function of halo mass for the {\sc Dark Sage} cosmology.

The radial profile Eq.~\ref{NFWden} gives the radial distribution of matter within halos, but we also need the radial distribution of galaxies, and these two may not be the same. In the one-halo term of the two-point correlation function given in Eq.~\ref{eq:haloone}, the radial distribution is expressed as the halo center-to-satellite density function $f_{cs}(x)$, and the satellite-to-satellite radial density function $f_{ss}(x)$. If we assume an NFW profile, and set $\alpha=2$ \citep{Berlind2002,Tinker2005,Zheng2007}, then $\rho_m(r)$ in Eq.~\ref{fcseq} is given by Eq.~\ref{NFWden}. The expression for $f_{ss}(x)$ is given in appendix A of \citet{Zheng2007}, but is lengthy and so not repeated here.

Plugging Eq.~\ref{NFWden} and \ref{NFWdencdcd} into Eq.~\ref{ygeq}, one can obtain the Fourier transformation of the NFW density, analytically expressed as \citep{Cooray2002}
\be \label{ygnfw}
\begin{split}
y_g(k)&=\frac{1}{\ln(1+c)-c/(1+c)}  \times \bigg \{ -\frac{\sin(ckR_s)}{(1+c)kR_s} \\
& +\sin(kR_s)\left[SI([1+c]kR_s)-SI(kR_s)  \right]   \\
&  +\cos(kR_s)\left[CI([1+c]kR_s)-CI(kR_s)  \right] 	\bigg \} ~,
\end{split}
\ee 
 and where
\begin{subequations}
\be 
SI(x) \equiv \int_0^x\frac{\sin t}{t}dt~,
\ee
\be
CI(x) \equiv -\int_x^{\infty}\frac{\cos t}{t}dt~,
\ee 
\end{subequations}
are the so-called Trigonometric integrals. This expression is used to calculate $y_g(k)$ of the NFW profile in Fig.~\ref{ygk}.

\section{The \HI--halo mass relation}\label{sec:hihmr}

\begin{figure} 
\centering
\includegraphics[width=\columnwidth]{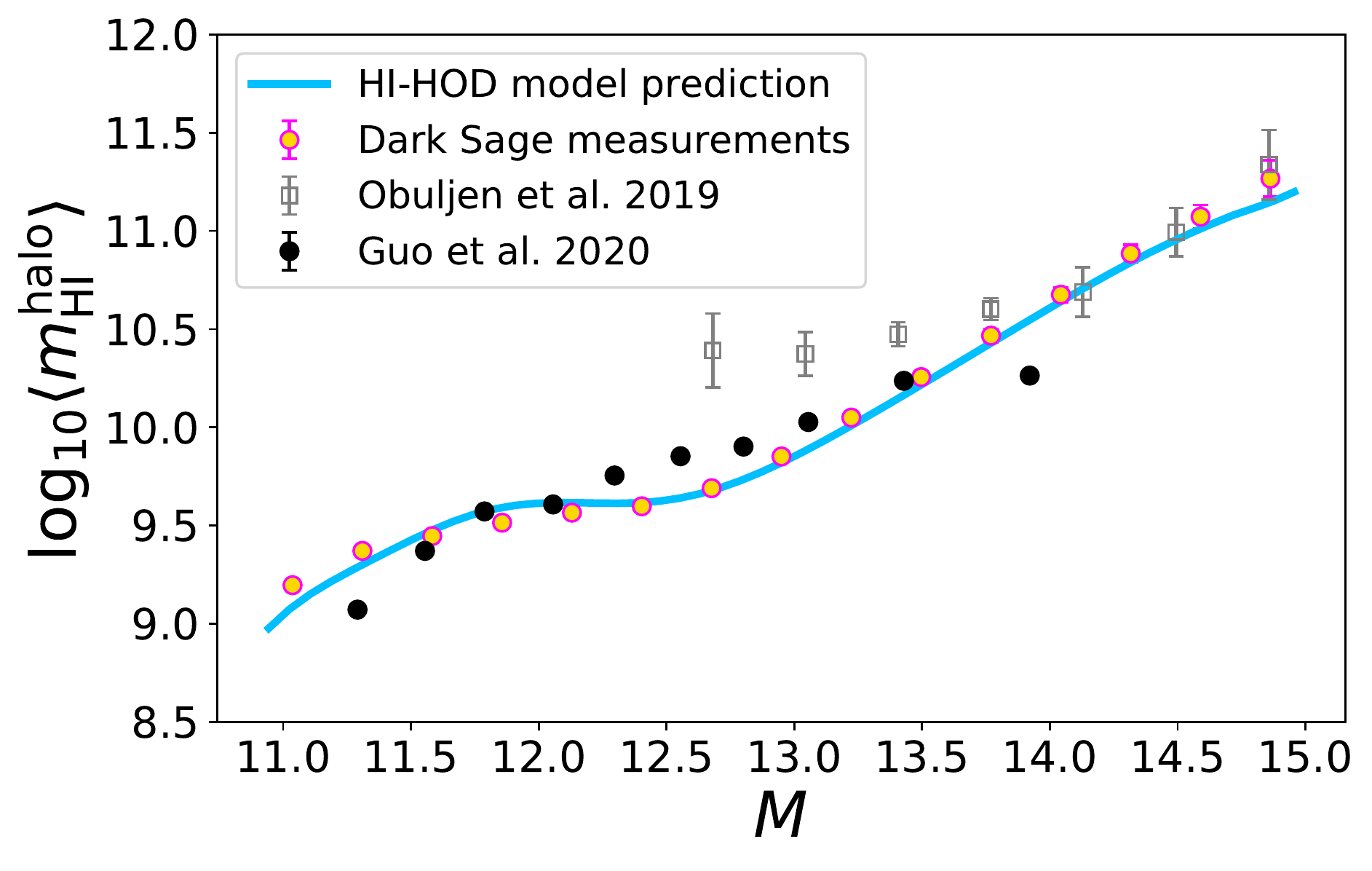}
 \caption{ { The mean \HI~content of halos (inside $2 R_{\rm vir}$) as a function of virial mass. Compared are the direct output of {\sc Dark Sage}, the prediction from our \HI-conditional HOD (fitted to {\sc Dark Sage} data, but not this relation specifically), and two observational datasets based on the ALFALFA survey \citep{Obuljen2019,Guo2020}. } }
 \label{HIHMR_HODs1}
\end{figure}

A predicted quantity of our \HI~HOD that we did not assess in the main text is the \HI--halo mass relation (HIHMR).  The HIHMR describes the \emph{total} \HI~content of a halo; that is, the summed \HI~content of all galaxies in a halo.  In principle, this should also include \HI~in the circumgalactic/intracluster medium \citep[see, e.g., the results of][]{Stevens2019,Voort2019,Garratt2021}, but here we assume this to be negligible.
In Fig.~\ref{HIHMR_HODs1}, 
we compare the \emph{mean} HIHMR from {\sc Dark Sage} and our HOD fitted to it with that derived from observations.

The yellow filled circles show the HIHMR measured from the direct output of {\sc Dark Sage}. The blue curve is the HIHMR predicted by the best-fit conditional HOD (with parameters given in the last two columns of Table~\ref{bkflb}). Reassuringly, 
we find good agreement between the measurements and prediction.

The gray squares and the black points display the HIHMR data from \citet{Obuljen2019} and \citet{Guo2020}, respectively. \citet{Obuljen2019} use the $\alpha.40$ sample of ALFALFA, while \citet{Guo2020} use the ALFALFA catalogue from \citet{Haynes2018}. They estimate the \HI ~masses in dark matter halos from galaxy groups. \citet{Guo2020} used $2R_{\rm vir}$ with $\Delta=200$ as the aperture for groups, while \citet{Obuljen2019} used $2R_{\rm vir}$ with $\Delta=180$ as their aperture (cf.~Eq.~\ref{Rvirs}).
 
At higher halo masses,  we find good 
agreement between our measurements and the observed values from \citet{Obuljen2019}.
And our measurements are generally in reasonable agreement with the observed values from \citet{Guo2020}.   
At lower halo masses, our measurements are slightly higher than the values from \citet{Guo2020}. Due to the observation limit of faint galaxies, the less massive galaxies of smaller groups might be missed in \citet{Guo2020},  resulting in the 
underestimated HIHMR in smaller halo mass bins (see Section 4.2 of their paper for more discussion). For a fair comparison to \citet{Obuljen2019} and \citet{Guo2020}, we removed the satellites that are beyond $2R_{\rm vir}$ to measure the HIHMR. For a more in-depth comparison method to the \citet{Guo2020} data for a semi-analytic model, see \citet{Chauhan2021}.


\bibliography{FQinRef}{}
\bibliographystyle{aasjournal}



\end{document}